\documentclass[matbio]{svjour}
\usepackage{amsmath,amssymb,graphics,epsfig,color,times,exscale,graphicx}
\usepackage{bbm,epic}

\input{psfig}

\newcommand\bs[1]{\boldsymbol{#1}}
\newcommand\mc[1]{\mathcal{#1}}
\newcommand\mf[1]{\mathfrak{#1}}

\newcommand{\RR}{\mathbb R}

\newcommand{\EE}{\mathbb E}
\newcommand{\QQ}{\mathbb Q}
\newcommand{\ZZ}{\mathbb Z}
\newcommand{\NN}{\mathbb N}

\newcommand{\cl}{c^{}_{\ell}}

\newcommand{\cz}{c_z}
\newcommand{\ch}{c_h}
\newcommand{\g}{g}
\newcommand{\h}{h}
\newcommand{\E}{E}
\newcommand{\dd}{{\, \rm d}}

\newcommand{\f}{g}
\newcommand{\F}{F}
\newcommand{\Phim}{\Phi_m}

\newcommand{\M}{M}
\newcommand{\hM}{\hat M}
\newcommand{\tH}{\tilde{H}}
\newcommand{\Q}{Q}
\newcommand{\U}{U}
\newcommand{\p}{p}
\newcommand{\w}{w}
\newcommand{\x}{x}
\newcommand{\y}{y}
\newcommand{\Z}{Z}
\newcommand{\z}{z}

\newcommand{\bF}{F}
\newcommand{\bM}{M}
\newcommand{\bH}{H}
\newcommand{\btH}{\tilde {H}}
\newcommand{\bPi}{\varPi}
\newcommand{\bQ}{Q}
\newcommand{\bR}{R}
\newcommand{\ba}{a}
\newcommand{\bp}{p}
\newcommand{\bpi}{\pi}
\newcommand{\tpi}{\tilde{\pi}}

\newcommand{\bh}{h}
\newcommand{\bv}{v}
\newcommand{\bw}{w}
\newcommand{\by}{y}
\newcommand{\bz}{z}

\newcommand{\1}{\bs{1}}

\newcommand{\cH}{{\mc{H}}}
\newcommand{\cM}{{\mc{M}}}
\newcommand{\cR}{{\mc{R}}}
\newcommand{\fS}{{\mf{S}}}
\newcommand{\cO}{{\mc{O}}}

\newcommand{\cP}{{\mc{P}}}
\newcommand{\fm}{{\mf{m}}}

\newcommand{\G}{\Gamma}

\newcommand\lambdamax{\lambda_{\rm max}}
\def\parspace{\cP_{(\Lambda_1,\dots,\Lambda_K)}}
\def\mutseq{\cM}
\def\stoma{\M}
\def\braseq{\cR}

\def\magn{b}

\def\macpar{\fm}
\def\ov{\omega}

\allowdisplaybreaks

\newcommand\ts{\hspace{0.5pt}}

\newcommand\Eref[1]{Equation~\textup{(\ref{eq:#1})}}
\newcommand\eref[1]{\textup{(\ref{eq:#1})}}

\newcommand\sref[1]{Section~\textup{\ref{sec:#1}}}

\makeatletter
\newcommand\cf{cf.\@ifnextchar,{}{\ }}
\newcommand\eg{e.g.\@ifnextchar,{}{, }}
\newcommand\ie{i.e.\@ifnextchar,{}{, }}
\makeatother

\let\oldcite\cite
\def\cite{\upshape\oldcite}
\smartqed
\makeatletter
\newif\if@showqed
\@showqedtrue
\global\@namedef{endproof}{\if@showqed\qed\fi\global\@showqedtrue\@endtheorem}
\makeatother
\makeatletter
\newlength\qedraise
\newcommand\qedhere{\@ifnextchar[{\qed@here}{\qed@here[0pt]}%] bracket matching
}
\def\qed@here[#1]{%
  \global\setlength{\qedraise}{#1}%
  {\@xp\aftergroup\csname\@currenvir @qed\endcsname}\global\@showqedfalse}%
  \def\displaymath@qed{%
    \relax
    \ifmmode
      \ifinner \aftergroup\linebox@qed
      \else
        \eqno
        \let\eqno\relax \let\leqno\relax \let\veqno\relax
        \raisebox{\qedraise}{\qed}%
      \fi
    \else
       \aftergroup\linebox@qed
    \fi
  }
  \@xp\let\csname equation*@qed\endcsname\displaymath@qed
  \@xp\let\csname multline*@qed\endcsname\displaymath@qed
  \def\align@qed{\tag*{\raisebox{\qedraise}{\qed}}}
  \@xp\let\csname align*@qed\endcsname\align@qed
\def\linebox@qed{\hfil\hbox{\qed}\hfilneg}
\makeatother

\begin{document}

\bibliographystyle{abbrv}

% for svjour
\title{An asymptotic maximum principle \\ for essentially
       linear evolution models}
\author{Ellen Baake \inst{1} \and Michael Baake \inst{2}
  \and 
  Anton Bovier \inst{3} \and Markus Klein \inst{4}}
\institute{
  Institut f\"ur Mathematik, Universit\"at Wien, Nordbergstr.\ 15,
  1090 Wien, Austria,
  \email{ellen.baake@univie.ac.at} 
  \and
  Fakult\"at f\"ur Mathematik,
  Universit\"at Bielefeld, Postfach 100131, 33501 Bielefeld, Germany, 
  \email{mbaake@mathematik.uni-bielefeld.de}
  \and
  Weierstrass-Institut f\"ur Angewandte Analysis und Stochastik,
  Mohrenstr.\ 39, 10117 Berlin, Germany, 
  and Institut f\"ur Mathematik, Technische Universit\"at Berlin,
  Stra\ss e des 17. Juni 136, 10623 Berlin, Germany,
  \email{bovier@wias-berlin.de}
  \and
  Institut f\"ur Mathematik, Universit\"at Potsdam, Postfach 60 15 53,
  14415 Potsdam, Germany, \email{mklein@math.uni-potsdam.de}} 
\titlerunning{Asymptotic maximum principle}
\authorrunning{Baake et al.}
\makeatletter
\def\@date{}
\makeatother
\def\copyleft{}
\def\makeheadbox{{%
\hbox to0pt{\vbox{\baselineskip=10dd\hrule\hbox
to\hsize{\vrule\kern3pt\vbox{\kern3pt
\hbox{\bfseries DRAFT}
\hbox{\today}
\kern3pt}\hfil\kern3pt\vrule}\hrule}%
\hss}}}
\maketitle

\begin{abstract}
Recent work on mutation-selection models has revealed that,
under specific assumptions on the fitness function and the
mutation rates, asymptotic estimates for the leading
eigenvalue of the mutation-reproduction matrix
may be obtained through a low-dimensional maximum principle
in the limit $N \to \infty$ (where $N$ is the number of types).
In order to extend this variational principle to a larger
class of models, we consider here a  family of
reversible $N \times N$ matrices and identify conditions under which
the high-dimensional Rayleigh-Ritz variational problem may
be reduced to a low-dimensional one that yields
the leading eigenvalue up to an error term of order $1/N$.
For  a large class of mutation-selection models, this implies estimates
for the mean fitness, as well as a concentration result for the
ancestral distribution of types.
\end{abstract}

\section{Introduction}
\label{sec:intro}

Many systems of population biology or reaction kinetics
may be cast into a form where individuals (or particles) of different types
reproduce and change type independently 
of each other in continuous time. If the types come from a
finite set $S$ and the population is so large that
random fluctuations may be neglected, one is led to
a linear system of differential equations of the form 
\begin{equation}\label{eq:lindgl}
  \dot \by = \by \bH
\end{equation}
with initial condition $\by(0)$. Here, $\by = (\y_i)_{i\in S}\in 
\RR_{\geqslant 0}^{|S|}$ holds the abundance  of the
various types; $\bH=(H_{ij})_{i,j\in S}$ is an 
$|S| \! \times \! |S|$ matrix, which represents a linear operator on
$\RR^{|S|}$. 
The main application we have in mind here is in population genetics,
where types are alleles, so that \Eref{lindgl} is
a haploid mutation-reproduction model; but one may
also think of a compartment model, where types are locations
of a certain chemical.  Important examples
of the analogous discrete-time dynamics include models of age-structured
populations, which are often referred to as matrix population models,
see Caswell's monograph \cite{Casw00}. In line
with large parts of the population genetics, and most of the
stochastics, literature, we will use the 
convention that $\by$ is a
row vector to which $\bH$ is applied from the right, so that
$H_{ij}$ ($i \neq j$) is the coefficient for the change from $i$ 
to $j$.

We will assume throughout that the linear operator
$H$ generates a positive semigroup, $\{\exp(tH) \mid t \geqslant 0\}$.
Since $S$ is finite,
this is equivalent to $H_{ij} \geqslant 0$ for 
$i\neq j$. The flow so generated leaves $\RR_{\geqslant 0}^{|S|}$
invariant.
We will further assume  that $\bH$ is irreducible 
(i.e., if $G(\bH)$ is
the directed graph with an edge from $i$ to $j$ if $i \neq j$ and
$H_{ij}>0$, then
there is a directed path from any vertex to any other vertex).

We will often use the decomposition
\begin{equation}\label{eq:decomp}
  \bH = \bM + \bR\,
\end{equation}
into a Markov generator $\bM$ and a diagonal matrix $\bR$. 
More precisely, we have $\bM=(M_{ij})_{i,j \in S}$ with $M_{ij}:=H_{ij}$
for $i \neq j$,  $M_{ii} := - \sum_{j \in S \setminus \{i\}} M_{ij}$
(so that $\sum_{j \in S} M_{ij}=0$), and
$\bR=\text{diag}\{R_i \mid i \in S\}$ with $R_{i} := H_{ii} - M_{ii}$.
Clearly, the decomposition in \eref{decomp} is unique, and $\bM$
is irreducible iff $\bH$ is, because $G(\bM)=G(\bH)$.
$\M_{ij}$ is the 
rate at which an $i$-individual produces $j$-offspring ($j \neq i$),
and $R_i$ is the net  rate at which
individuals of type $i$ reproduce themselves; this may also include 
death terms and thus be negative.

Solutions of \eref{lindgl} cannot vanish altogether (unless
$\by(0)=0$), since 
$\text{tr}(H)$ is finite, hence $\det \big (\exp(tH) \big ) = 
\exp(t \, \text{tr}(H)) > 0$ and $\text{ker} \big (\exp(tH) \big ) = \{0\}$,
for all $t \geqslant 0$.
Therefore, we
may also consider the corresponding normalized equation for the 
proportions $\p_i := \y_i / (\sum_{j \in S} \y_j)$, which is 
often more
relevant. Clearly,
\begin{equation}\label{eq:nonlindgl}
  \dot p_i = \sum_{j \in S} p_j M_{ji}  
           + \big ( R_i - \sum_{j \in S} R_j p_j \big )  p_i\,.
\end{equation}
In the population genetics context, this is the  mutation-selection
equation for a haploid population, or a diploid one without
dominance; for a comprehensive review of this class
of models, see \cite{Buer00}.
It is well known, and easy to verify, that 
the way back from \eref{nonlindgl}
to \eref{lindgl} is achieved through the transformation \cite{TMcB74}
\[
   \by(t) := 
   \bp(t) \exp \Big ( \sum_{j \in S} R_j \int_0^t  p_j(\tau)  d\tau \Big )\,. 
\]
This substitution can thus be viewed as a global linearization transformation
and explains why \eref{nonlindgl} is an `essentially linear' equation.
In fact, Eq.~\eref{nonlindgl} appears in a variety of contexts. In particular,
its discrete-time relative may be used to describe the dynamics of 
the age structure of a  population, compare \cite[Ch.\ 4]{Bulm94}.
Due to its frequent appearance, a better understanding of 
Eq.~\eref{nonlindgl} and its solutions is the main motivation for 
the present work. 

Clearly, the solution of \eref{nonlindgl} is obtained from that
of \eref{lindgl} through normalization:
\[
  \by(t) = \by(0) \exp(t\bH), \quad \bp(t) = \frac{\by(t)}{\sum_i \y_i(t)}\,.
\]
Of course, proportions of types in a population that grows
without restriction (which is biologically reasonable only over short
time scales) do not represent the only way in which \eref{nonlindgl} may
arise. Actually, the same equation for 
$\bp$ results if \eref{lindgl} is replaced by 
\[
   \dot \by = \by \big (\bH - \gamma(t) \big ) \,,
\]
where  $\gamma(t)$ is some scalar (possibly nonlinear) function
which describes the elimination of individuals by population 
regulation. This is obvious from the invariance of \eref{nonlindgl}
under $R_i \to R_i + \gamma(t)$ if performed simultaneously for
all $i$. The function $\gamma(t)$ may, for example, describe 
an additional death term caused by crowding,
which may depend on $t$ through $y$, but acts on all types
in the same way.

Eq.\ \eref{nonlindgl} may be read in two ways (cf.\ \cite{Hofb85}).
If mutation and reproduction go on independently of each other, the 
parallel (or
decoupled) version is adequate. Here, every
$i$-individual gives birth to offspring of its own type at rate
$B_i$, dies at rate $D_i$, and mutates to $j$ at rate $M_{ij}$
($j \neq i$). Then $R_i := B_i - D_i$  is the net reproduction rate
or Malthusian fitness \cite[Ch.\ 5.3]{CrKi70}, and Eq.\ \eref{nonlindgl} is
immediate. If, however, mutation is a side effect of reproduction
(through copying errors of the replication process, for example),
the coupled version  is more relevant \cite{Akin79,Hade81}.
When an $i$-individual
reproduces (which it does, as before, at rate $B_i$, while it dies
at rate $D_i$), the offspring is
of type $j$ with probability $V_{ij}$ ($\sum_j V_{ij} = 1$).
This leads to
\begin{equation}\label{eq:coupled}
  \dot p_i = \Big ( \sum_{j \in S} p_j B_j V_{ji} \Big ) 
           - \Big ( D_i + \sum_{j \in S} R_j p_j  \Big ) p_i\,,  
\end{equation}
where, again, $R_i = B_i - D_i$. But if we  set 
$M_{ij} := B_i (V_{ij}-\delta_{ij})$,
we arrive again at Eq.~\eref{nonlindgl}. 
In both cases, $\sum_j R_j p_j$ is the mean fitness of
the population. Obviously, a mixture of both the parallel and the coupled
mutation mechanisms can be tackled in a similar way. 
Furthermore, the decoupled model arises as the weak-selection
weak-mutation limit of the coupled one \cite{Hofb85}, or of
the corresponding model in discrete time \cite[p.\ 98]{Buer00}.

The model \eref{coupled} also arises in the infinite population limit 
of the well-known Moran model with selection and mutation,
see \cite[Ch.\ 3]{Ewe04} or \cite[p.\ 126]{Durr02}. 
This is a {\em stochastic} model where, in a population of $m$ 
individuals, every individual of type $i$ reproduces at rate $B_i$, 
and the offspring, which is of type $j$ with probability $V_{ij}$,
replaces a randomly chosen individual in the population 
(possibly its own parent). 
To describe the entire population, let $Z_i(t)$ be the random variable that
gives the number of $i$-individuals at time $t$, and 
$Z(t)=\big (Z_i(t) \big)_{i\in S}$.  Hence, if $Z(t)=z$, and $j \neq k$,
we can have transitions from $z$  to $z + e_j - e_k$,
where $e^{}_j$  denotes the unit vector corresponding to $j$.
Such a transition occurs at rate 
$\sum_i B_i V_{ij} z_i z_k / m$. 
Let us look at the influence of increasing $m$, whence we write
$Z^{(m)}(t)$ to indicate dependence on system size.
As $m \to \infty$, the sequence of random processes
$Z^{(m)}(t)/m$ converges  
pointwise almost surely, and even uniformly for every 
finite interval $[0,t]$, to the solution of
the differential equation \eref{coupled} with $D_i \equiv 0$,
and initial condition $Z^{(m)}(0)/m$ (resp.\ its limit as $m\to\infty$), 
compare \cite[Thm.\ 11.2.1]{EtKu86}.

The linear equation \eref{lindgl}  has a more direct stochastic
interpretation in terms of a continuous-time multitype branching
process. After an exponential waiting time with
expectation $\tau_i$, an individual of type $i$ produces a random offspring
with a finite expectation of $b_{ij}$ children of type $j$
(we will not specify the distribution explicitly since we will
not fully develop the stochastic picture here). The matrix
$\bH$ with $H_{ij} =  (b_{ij} - \delta_{ij})/\tau_i$ 
then is the generator 
of the first-moment matrix. That is, if
$\Z_j(t)$ is again the (random) number of individuals of type $j$
at time $t$, and $\EE^i$ the associated 
expectation in a population started by a single $i$ individual at time $0$, 
then
\begin{equation}\label{eq:expect}
  \EE^i(\Z_j(t)) = \big ( \exp(t\bH) \big )_{ij}.
\end{equation}
Furthermore, with the identification
$\y_i(t) = \EE \big ( \Z_i(t) \big )$,
\Eref{lindgl} then simply is the forward equation for
the expectations.  (See \cite{AtNe72} or \cite{KaTa81} for the
general context of multitype branching processes, and 
\cite{HRWB02} for the application to mutation-selection models.)

Important first questions concern the asymptotic properties of
the systems discussed. A key to these properties is the 
leading eigenvalue, $\lambdamax$, of $\bH$
(i.e., the real eigenvalue exceeding the real parts of all 
other eigenvalues).
If, on short time scales, unrestricted
growth according to \eref{lindgl} is relevant, then 
$\lambdamax$ is the asymptotic growth rate of the population
(and is related to the chance of ultimate survival). 
The stationary distribution of types in \eref{nonlindgl} is given by
the left eigenvector of $\bH$ corresponding to $\lambdamax$.
We will call it the {\em present distribution} of types, as opposed to the
(less well-known, but equally important) {\em
ancestral distribution} that is obtained by picking individuals
from the present distribution and following their ancestry
backward in time until a new stationary state is reached.
This ancestral distribution
is given by the elementwise product of the left and right
eigenvectors of $\bH$ corresponding to $\lambdamax$, with proper
normalization \cite{Jage89,Jage92}. 
The knowledge of $\lambdamax$ is a prerequisite for the
calculation of these eigenvectors.
In the population genetics context, the
present distribution is often referred to as mutation-selection balance, with
$\lambdamax$ as the mean fitness.  Finally, and perhaps most importantly,
the dependence of $\lambdamax$ on certain model parameters
is of great interest. For example, a lot of research
has been directed towards the question of how the mean fitness
changes when the mutation rate increases (i.e., when
$\bM$  is varied by some nonnegative scalar factor), and
interesting effects have been observed, for example so-called error thresholds.
They may be defined as non-analytical changes of $\lambdamax$ that
occur when the mutation rate surpasses a critical value,
in analogy
with a phase transition in physics. This is accompanied  by
a discontinuous change in the ancestral distribution, as well as
pronounced
changes in the present distribution of types;
see \cite[Ch.\ III]{Buer00} and \cite{EMcCS89} for general reviews,
\cite{HRWB02} for  recent results and a classification of the 
various  threshold phenomena that may occur, and \cite{GeHw02} for a
recent application to the evolution of regulatory DNA motifs.

In general, exact expressions for eigenvalues are hard to obtain 
if $\lvert S \rvert$ is large but fixed.
In recent work on mutation-selection models, however,  scalar or
low-dimensional
maximum principles for the leading eigenvalue have been identified
for certain examples  in a suitable continuous limit as 
$\lvert S \rvert \to \infty$, see \cite{HRWB02,GaGr04}.
It is the purpose of this paper to generalize these results to a
large class of operators. We will do so under the general
assumption that the Markov generator $M$ is reversible,
which means that the equilibrium  flux
from state $i$ to state $j$ is the same as that from $j$ to $i$.
This entails that the mutation process is the same in the forward
and backward direction of time, and 
covers many of the  frequently-used  models in classical
population genetics, 
for example, the house-of-cards model, and the random-walk
mutation model with Gaussian mutant distribution 
(see \cite[Ch.\ 3]{Buer00} for its definition, \cite{Redn04}
for the reversibility aspect, and a more general class of
reversible random-walk mutation models).  
Also, practically all models of nucleotide evolution that are in use in
molecular population genetics, like the Jukes-Cantor, Kimura,
Felsenstein, and HKY models, cf.\ \cite{SOWH95} or \cite[Ch.\ 13]{EwGr01},
are reversible.
This property is particularly important in phylogenetic inference,
where one relies on looking back from the present into the past.

The paper is organized as follows. In Section \ref{sec:rev},
we will apply the Rayleigh-Ritz maximum principle to our class
of matrices. This leads to a high-dimensional problem, which is
hard to solve in practice. An example of how the problem may be reduced
to a  scalar one is given in Section \ref{sec:ex}. The main results are
presented in Section \ref{sec:maxeval}. Here, we identify fairly 
general conditions under which the high-dimensional problem may
be reduced to a low-dimensional variational problem that yields
the leading eigenvalue up to an error term of order $1/N$, in the limit
$N=\lvert S \vert \to \infty$. 
Sections \ref{sec:lump} and \ref{sec:seqspace} are devoted to
the lumping procedure. 
They  show that a large class of  models 
on a type space $S$ arises, in a natural  way, from 
models defined on a `larger' space $\fS$, by combining several types
in $\fS$ into a single one in $S$. The general framework is set out
in Section \ref{sec:lump}, and in Section \ref{sec:seqspace},
we apply it to the important case
where $\fS$ is the space of all sequences of fixed length 
over a given alphabet. Section \ref{sec:applic} makes the
connection back to the maximum principle and shows how the
lumping procedure may lead to `effective' models (on $S$) to
which our asymptotic results may then be applied. The Hopfield
fitness function, along with sequence space mutation, emerges
as an example.  In Setion 8, we summarize our findings
and discuss them informally, and in a more biological context.

\section{The general maximum principle for reversible generators}
\label{sec:rev}
Let us first fix our assumptions and notation.
Since we assume $\bM$ to be an irreducible Markov generator,
Perron-Frobenius theory, cf.\ \cite[Appendix]{KaTa75}, tells us that it
has a leading eigenvalue $0$
which exceeds the real parts of all other eigenvalues,
and an associated strictly positive left 
eigenvector $\bpi$. This will be normalized s.t.\ 
$\sum_i \pi_i=1$; 
then, $\bpi$ is the stationary distribution of the Markov semigroup
generated by $\bM$.

We will assume  that $\bM$ is
reversible, i.e., 
\begin{equation}\label{eq:rev}
 \pi_i \M_{ij} = \pi_j \M_{ji}
\end{equation}
for all $i$ and $j$,
which also entails $\pi_i H_{ij} = \pi_j H_{ji}$ since $\bR$ is
diagonal.
Likewise, due to irreducibility, the leading eigenvalue, $\lambdamax$,
of $\bH$ is simple; we will encounter the corresponding eigenvectors
in due course.

Let  us note in passing that, due to 
reversibility combined with
irreducibility, the equilibrium
distribution $\pi$ of $\bM$ is
available explicitly as follows \cite[p.~35]{Keil79}.
Let $(v_1,v_2,\dots,v_{|S|})$
be the vertices of the directed graph $G(\bM)$ 
(with $(v_i,v_j)$ a directed edge iff $M_{ij} > 0)$.
Since  $\pi_i > 0$ for all $i \in S$, $(v_j,v_i)$ is an edge iff $(v_i,v_j)$
is, as a consequence of \eref{rev}. 
Now, set $\tpi_{1}=1$ and consider any
$2\leqslant \ell \leqslant |S|$. By irreducibility, there is a directed
path along $v_1 = v_{k_0}, v_{k_1}, \ldots, v_{k_m} = v_{\ell}$, which
also exists as a path in reverse direction. If we now set
\begin{equation}\label{eq:pi} 
  \tpi_{k_{\ell}} 
  = \prod_{j=1}^{m} \frac{M_{k_{j-1},k_{j}}}{M_{k_{j},k_{j -1}}}\,,
\end{equation}
$\pi_i = \tpi_i / (\sum_{j \in S} \tpi_j)$
is the stationary probability distribution
of the Markov generator $\bM$. This reflects the path
independence of reversible Markov chains \cite[p.~35]{Keil79}:
For any path  with an arbitrary number $m+1$ of vertices 
$(k_0,k_1,\dots,k_{m})$ in our graph $G(\bM)$, 
the product  
$\prod_{j=1}^{m} (M_{k_{j-1},k_{j}} / M_{k_{j},k_{j -1}})$
only depends on the initial and final vertices, $k_0$ and $k_m$,
not on the path connecting them. 
Note that, if $G(\bM)$ admits a Hamiltonian path, the calculation
in \eref{pi} can be further simplified by following such a path
edge by edge.

It  is well-known that reversibility has 
important consequences
for eigenvalues and eigenvectors of a Markov generator. An excellent
exposition for the closely-related discrete-time case is Chapter~2.1
of \cite{Brem99}. Following these lines, we now define,   
for $i \neq j$, 
\begin{equation} \label{eq:Fij}
   \F_{ij} \, := \, \sqrt{\pi_i} \, \M_{ij} \, \frac{1}{\sqrt{\pi_j}}
   \, = \, \F_{ji}\,,
\end{equation}
where the symmetry follows from the reversibility of $\bM$.
Clearly, $\F_{ij} \geqslant 0$ and 
$\F_{ij} = (\F_{ij}\F_{ji})^{1/2} = (\M_{ij}\M_{ji})^{1/2}$.
As a consequence, the matrix 
\begin{equation}
  \btH := \bPi^{1/2} \bH \bPi^{-1/2}
\end{equation}
with $\bPi := \text{diag} \{\pi_i \, | \, i \in S\}$ 
has off-diagonal entries $F_{ij}$, is   symmetric and  has
real spectrum identical to that of  $\bH$, with correspondingly transformed
eigenvectors.
We now decompose  $\btH$ in the same way as we did with $\bH$ in
\eref{decomp}, namely into a Markov generator $\bF$
plus a diagonal matrix $\E$.
To this end, let $\bF= (\F_{ij})_{i,j \in S}$ with $\F_{ij}$ as in
\eref{Fij} for $i \neq j$, and complete this by
$\F_{ii} := - \sum_{j \in S \setminus \{i\}} \F_{ij}$. With 
\[
  \E_i := R_i + \M_{ii} - \F_{ii} 
  = R_i + \sum_{\substack{j \in S \\ j > i}} \big ( 2 \sqrt{M_{ij} M_{ji}} 
    - (M_{ij} + M_{ji}) \big ),
\]
one now has
$\tH_{ij} = \F_{ij} + \E_i \delta_{ij}$ for all $i,j \in S$, i.e.,
\begin{equation}\label{eq:decomp2}
  \tH = \F + \E
\end{equation}
with $\F$ a Markov generator and $\E=\text{diag}\{\E_i \mid i \in S\}$.

This now allows us to formulate a suitable variant of the 
Rayleigh-Ritz (or Courant-Fisher) maximum principle 
for the leading eigenvalue of
$\btH$, compare \cite[Thm.\ 19.4]{Praso96}. Clearly,
\begin{eqnarray}\nonumber
  \lambdamax & = & \sup_{\bv: \sum_{\ell \in S} v_{\ell}^2 = 1} \,
                   \sum_{i,j \in S} v_i \tH_{ij} v_j\\ 
             \label{eq:RR}
             & = & \sup_{\bv: \sum_{\ell \in S} v_{\ell}^2 = 1}
                   \Big ( \sum_{i, j \in S} v_i \F_{ij} v_j
                   + \sum_{k \in S} \E^{}_k  v_k^2 \Big )\,,
\end{eqnarray}
where we have used the decomposition \eref{decomp2} in the second
step. Note that the supremum is, indeed, assumed, since the
space of probability measures on $S$ is compact.
The maximizer, i.e., the normalized principal  eigenvector 
of $\btH$, is unique and strictly
positive (since the same holds for the corresponding eigenvector of $\bH$),
so that the above may also be read as an $L^1$ variant through the substitution
$\nu_i := v_i^2$. 

Note that,
since $\bF$ is a Markov generator, the quadratic form 
$\sum_{i, j \in S} v_i \F_{ij} v_j$ is negative
semidefinite with maximum $0$, which is assumed for the stationary
distribution of  $\bF$ 
(since $\bF$ is symmetric and irreducible, this is the equidistribution,
and unique). We thus have a simple upper bound on $\lambdamax$:
\begin{equation}\label{eq:upbound}
  \lambdamax \leqslant \sup_{\bv: \sum_{\ell \in S} v_{\ell}^2 = 1} \;
                       \sum_{k\in S} \E_k^{}  v_k^2 
  = \max_{k \in S} \E_k\,,
\end{equation}
while we can obtain a lower bound for any $v\geqslant 0$ with 
$\sum_{\ell} v_{\ell}^2=1$ via
\begin{equation}\label{eq:lobound}
    \sum_{i, j \in S} v_i \F_{ij} v_j + \sum_{k \in S} \E_k^{}  v_k^2
    \leqslant \lambdamax\,.
\end{equation}

Even though each step of the above derivation is elementary, it
is worthwhile to summarize the findings as follows.

\begin{proposition}\label{prop1}
Let $S$ be a finite set, and let $\bH$ be an 
$\lvert S \rvert \! \times \! \lvert S \rvert$-matrix
with decomposition $\bH=\bM+\bR$ into an irreducible and reversible
Markov generator $\bM$ and a diagonal matrix $\bR$. If $\bpi$ is the
stationary distribution of $\bM$, $\bH$ can be symmetrized to
$\btH=\Pi^{1/2}\bH\Pi^{-1/2}$ with $\Pi={\rm diag} \{\pi_i \mid i \in S\}$.
The matrices $\bH$ and $\btH$ are isospectral, and their leading 
eigenvalue $\lambdamax$ is given by the maximum principle 
\eref{RR}. Furthermore, simple upper and lower bounds for $\lambdamax$
are provided by Eqns.~\eref{upbound} and \/ \eref{lobound}.
\endproof
\end{proposition}
It is our aim to identify  conditions under which the inequality 
\eref{upbound} becomes an equality, at least asymptotically as
$\lvert S \rvert \to \infty$.

As a first step, consider the maximizer of \eref{RR}, i.e.,
the principal eigenvector $\bw$ of $\btH$, normalized via
$\sum_{i\in S} w_i^2=1$. Since $\btH$ is a symmetric matrix,
we have $ \bw \btH = \lambdamax \bw$ and, simultaneously,
$\btH \bw^T = \lambdamax \bw^T$. 
Hence, 
\begin{equation}\label{eq:hz}
  \bz^T := \cz \bPi^{-1/2} \bw^T \quad \text{and} \quad  
  \bh := \ch \bw \bPi^{1/2} 
\end{equation}
are the principal right and left
eigenvectors of $\bH=\bPi^{-1/2} \btH \bPi^{1/2}$. 
We will adjust the constants $\ch$  and $\cz$ s.t.\ 
$\sum_i \h_i=\sum_i \h_i \z_i = 1$; clearly, this
implies $\cz \cdot \ch =1$. 

The vector $\bh$   gives the stationary distribution of types 
in \Eref{nonlindgl}.
Furthermore, it is well-known that, for irreducible $\bH$ and
$t \to \infty$, the matrix 
$\exp(t(\bH -\lambdamax \1))$ becomes a projector
onto  $\bh$, with matrix elements
$\z_i \h_j$ (compare \cite[Appendix]{KaTa75}). 
Therefore,
\begin{equation}
     \lim_{t \to \infty} 
    \frac{\sum_{j\in S} \big (\exp{(t\bH})\big )_{ij}}{\sum_{k,\ell \in S} 
    \h_k \big ( \exp{(t\bH})\big )_{k \ell}} = 
    \frac{\sum_{j \in S} \z_i \h_j}{\sum_{\ell \in S} \h_{\ell}} =
    \z_i\,.
\end{equation}
With \eref{expect} in mind, $\z_i$ may therefore be understood as
the asymptotic offspring expectation of an $i$ individual,
relative to the mean offspring expectation of an equilibrium
population. If $\bR = C \1$ for some constant $C$,
we have $\z_i \equiv 1$, in line with the fact
that $\bH - C \1$ is then a Markov generator.

From \eref{hz}, along with the normalization of $\bh$ and $\bz$,
the relations
\begin{equation}\label{eq:hpiz}
   \h_i   =  \frac{\pi_i \z_i}{ \sum_{j \in S} \pi_j \z_j}
   \quad \text{and} \quad
   \w_i^2  =  \h_i \z_i
\end{equation}
are obvious. In particular, with
\begin{equation}\label{eq:a}
    a_i := \w_i^2 = \h_i \z_i > 0\,,
\end{equation}
we obtain the corresponding $L^1$-maximizer of \eref{RR}.

To arrive at another interpretation of $\ba$, consider  the Markov 
generator $\bQ$ with elements
\begin{equation}\label{eq:Qij1}
   \Q_{ij} = \z_i^{-1} (H_{ij} - \lambdamax \delta_{ij}) \z_j\,.
\end{equation}
It is easily confirmed that $\bQ$ is indeed a Markov
generator (i.e., $Q_{ij} \geqslant 0$ for $i \neq j$, and 
$\sum_j Q_{ij} = 0$).
Using \eref{hpiz} and reversibility, one
observes that $\bQ$ may also be written as
\begin{equation}\label{eq:Qij2}
  \Q_{ij} = \h_i^{-1} (H_{ji} - \lambdamax \delta_{ij}) \h_j\,.
\end{equation}

In this form, $\bQ$ is the generator of the 
backward process on the stationary distribution
as described in \cite[Corollary 1]{Jage92}  for general multitype branching
processes, and used in \cite{HRWB02} in the context of mutation-selection
models. Loosely  speaking, $\bQ$ describes the Markov chain 
which results from
picking individuals randomly from the stationary
distribution $\bh$ and following their lines of descent
backward in time. Eq.\ \eref{Qij1} is the
corresponding forward version as used in \cite{Jage89}
and  \cite{GeBa03}.
It is immediately verified that $\bQ$ has 
principal left eigenvector 
(i.e., stationary distribution) $\ba$. 
This  is  known as the {\em ancestral distribution} of types
(as mentioned in the Introduction); 
its properties are analyzed  in \cite{GeBa03}.
Let us  summarize this as follows.
\begin{proposition} \label{prop:a}
Let the assumptions be as in Proposition $1$. Then, the principal 
eigenvector $\bw$  of $\btH$ gives the principal left and right 
eigenvectors of $\bH$ and their mutual relations through
Eqns.~\eref{hz} and \eref{hpiz}. The $L^1$-maximizer $a=(a_i)_{i\in S}$
of \eref{RR} admits the interpretation of an ancestral distribution
as the stationary state of the backward Markov generator \/ $Q$ of
\eref{Qij1} and \eref{Qij2}.
\endproof
\end{proposition}

\section{A scalar maximum principle: An example}
\label{sec:ex}
The maximum principle \eref{RR} is not very useful in practice if
$|S|$ is large but fixed, since maximization is then over a large
space. In \cite{HRWB02}, this high-dimensional maximization could be
reduced to a scalar one for  special choices of $\bM$ and $\bR$. We will
re-derive this result here in a simplified way, which will also serve
as an introduction to the more general methods and results we are aiming at.
Let $S = \{0,1,\ldots,N\}$ with the following mutation scheme:
\begin{equation*}
  \boxed{0} \quad
  \substack{\xrightarrow{U_0^+} \\ \xleftarrow[U_1^-]{}}
  \quad \boxed{1} \quad
  \substack{\xrightarrow{U_1^+} \\ \xleftarrow[U_2^-]{}}
  \quad \boxed{2} \quad \cdots \quad
  \substack{\xrightarrow{\hspace{3.5pt} U_{i}^+ \hspace{3pt}} \\ 
  \xleftarrow[U_{i+1}^-]{}}
  \quad \cdots \quad
  \boxed{N\!\!-\!\!1} \quad
  \substack{\xrightarrow{U_{N\!-1}^+} \\ 
  \xleftarrow[\hspace{4pt} U_N^- \hspace{4pt}]{}}
  \quad \boxed{N}
\end{equation*}

Suppressing the  (relevant!) dependence on $N$ in the notation, we
then have
\begin{equation}
   \M_{i,i+1}  =  \U_i^+, \quad 
   \M_{i,i-1}  =  \U_i^- 
\end{equation}
for $i \in S$, where we set $\U_N^+=\U_0^-=0$.
This is a variant of the so-called single-step mutation model 
of population genetics
\cite[Ch.\ III.4]{Buer00}. It emerges if sequences of sites 
(nuceotide sites or
loci) are considered, and the `type' is identified with the
number of sites at which the sequence differs from a given
reference sequence or wildtype; see \cite{RWC03} for a
recent application. If fitness is a function of this
number only, and if mutations occur independently of
each other in continuous time, we are in the setting of the
single-step mutation model.

Hence, for all $i \in S$, we have
\begin{equation}
  \F_{i,i+1} = (\M_{i,i+1} \M_{i+1,i})^{1/2} = (\U_i^+\U_{i+1}^-)^{1/2}
             = \F_{i+1,i}
\end{equation}
with the obvious meaning for $i=0$ and $i=N$; also, $\F_{ij}:=0$
whenever either $i$ or $j$ is not in $S$, or if $\lvert i-j \rvert > 1$.
In order to evaluate the lower bound in \eref{lobound},
let $N$ be large,  $1 \leqslant L \ll N$, and $\ell \in S$. We
will use the simple test function 
$\nu := (\nu_0,\nu_1,\ldots,\nu_N)$ defined through 
\[
   \nu_i = \cl \cdot \begin{cases} 0, & 
           i \notin \big ( \ell + [-L,L]  \big ) \cap S\\
                                   1, & 
           i \in \big ( \ell + [-L,L] \big ) \cap S
                     \end{cases}
\]
with  $[-L,L] := \{-L,-L+1, \ldots, L-1, L\}$, and the
constant $\cl$ chosen so that $\sum_i \nu_i=1$. That is, 
$\nu$ is a normalized step function around $\ell$,
which does not extend beyond $0$ or $N$.
If $\ell + [-L,L] \subset S$, one always has
$\cl=1/(2L+1)$; a short calculation shows that, in any case,
\[
  \frac{1}{2L+1} \leqslant \cl \leqslant \frac{1}{L+1}\,,
\]
due to $L \ll N$.
With $\nu^{}_i = v_i^2$,
the quadratic form in \eref{RR} and \eref{lobound} reduces to
\[
 \sum_{i,j \in S} v_i \F_{ij} v_j = 
 \cl \sum_{i,j \in \ell+[-L,L]} \F_{ij} = 
 -\cl (\F_{\ell-L,\ell-L-1}+\F_{\ell+L,\ell+L+1})\,,
\]
due to the tridiagonal nature of the Markov generator $\F$.
Since
\[  
 \frac{1}{2}(\F_{\ell-L,\ell-L-1}+\F_{\ell+L,\ell+L+1}) 
 \leqslant  \max_{i \in S} \F_{i,i+1} = \max_{i,j \in S} \F_{ij} =: \F_{\rm max},
\]
one has
\begin{equation}\label{eq:qf}
   \Big | \sum_{i,j\in S} v_i \F_{ij} v_j \big | 
   \leqslant \frac{2 \F_{\max}}{L+1}\,.
\end{equation}
On the other hand, the second term in \eref{RR} resp.\ \eref{lobound}
(to be called the `diagonal part' in what follows) becomes
\begin{equation}\label{eq:diag}
  \sum_{i\in S} \E_i v_i^2 
  = \cl \sum_{i=\ell-L}^{\ell+L} \Big (R_i - \U_i^+ - \U_i^- 
  + \sqrt{\U_i^+ \U_{i+1}^-} + \sqrt{\U_i^- \U_{i-1}^+} \, \Big ) \,,
\end{equation}
where $\U_i^{\pm} := 0$ is implied whenever $i \notin S$.

Employing Landau's $\cO$-notation \cite[Ch.\ 1]{Brui81}, we now assume that
\begin{equation}\label{eq:approxur}
  \U_i^{\pm}  =  u^{\pm}(\x_i) + \cO(1/N)  \quad \text{and} \quad
   R_i        =  r(x_i) + \cO(1/N)
\end{equation}
with continuous functions $u^+$, $u^-$, and $r$ on [0,1], and 
the new `type variable' $\x_i = i/N$; it is further implied that
the constant in the $\cO(1/N)$ bound
is uniform for all $i$.
(Eq.\ \eref{approxur} differs   from the scaling in
\cite{HRWB02} by a global factor of $N$, which means nothing 
but a change of the time scale.)

Define $\g(x) := u^+(x) + u^-(x) - 2 \sqrt{u^+(x)u^-(x)}$,  let
$\x^*$ be a point  at which $r(x)-\g(x)$ assumes its supremum,
and choose $\ell := \lfloor N \x^* \rfloor$. With an appropriate scaling
of $L$ (such as $L \sim \sqrt{N}$, to be specific),
the right-hand side of \eref{qf} is $\cO(1/\sqrt{N})$.
In \eref{diag}, the sum has $\cO(\sqrt{N})$ terms,
which is balanced by $\cl=\cO(1/\sqrt{N})$; together with 
\eref{approxur}, this turns the right-hand side
of \eref{diag} into
$r(x^*)-\g(x^*)  + \cO(1/N)$. At the same time, the upper bound in
\eref{upbound} also behaves like $r(x^*)-\g(x^*)  + \cO(1/N)$.
Thus, the right-hand side of \eref{qf} contributes the largest
error term, so that we obtain the
asymptotic maximum principle 
\begin{equation}\label{eq:rg}
  \lambdamax = \sup_{\x \in [0,1]} \big ( r(x)-g(x) \big )
\end{equation}
up to $\cO(1/\sqrt{N})$, as $N \to \infty$. 

Finally, recall from \sref{rev} that, for finite $N$,
the maximizer of \eref{RR} is unique and  given by the ancestral distribution 
$a=(\h_i \z_i)_{i \in S}$. However, in the limit as $N \to \infty$,
uniqueness may be lost, which is also reflected by the fact that
the supremum in \eref{rg} may be assumed at more than one point.  It is
these degenerate situations where error thresholds may occur \cite{HRWB02}.

\begin{remark}
The maximum principle \eref{rg} also holds for
functions $r$ and $u^{\pm}$ with a finite number of jumps \cite{HRWB02}.
This can be dealt with  in the current framework with slightly
more effort, but we avoid this here to keep the example as
transparent as possible.
\end{remark}
\begin{remark}
With a more careful choice for the scaling of $L$, one gets the
quadratic form (defined by the matrix $F$)
down to $\cO(1/N^{1-\varepsilon})$ for arbitrary
$\varepsilon>0$, but $\cO(1/N)$ is only obtained with the help
of better (smooth) test functions. This will now be done.
\end{remark} 

\section{An asymptotic maximum principle: the general case}
\label{sec:maxeval}

The maximum principle allows for an asymptotic estimation of the 
leading eigenvalue when the Markov generator $F$ can be considered as 
`small' in a suitable sense, in comparison to the derived
effective `diagonal' part as defined by $E$.
Before stating precise conditions and results, 
let us briefly discuss the heuristics behind this. Due to the 
symmetry of $F$, we can rewrite Eq.~\eref{RR} as
\begin{equation}  \label{eq:RR.1}
  \lambdamax   
              =  \sup_{\bv: \sum_{\ell \in S} v_{\ell}^2 = 1}
                   \Big ( -\frac 12\sum_{i, j \in S}  \F_{ij} (v_i-v_j)^2
                   + \sum_{k \in S} \E^{}_k  v_k^2 \Big )\,.
\end{equation}
Thus, it is obvious that the $\F$-term  favours 
constant $v$ while the diagonal $\E$-part favours $v$ that are concentrated on 
the  points $k$ where $\E_k$ is maximal. 
Clearly, the outcome of this 
competition depends on some concentration and smoothness properties of the  
matrices involved. 

For simplicity, let us now assume that our set $S$ consists of integers
or, more generally, $d$-tuples of integers. So, $S \subset \ZZ^d$, with
$\lvert S \rvert < \infty$. 
(It will become apparent later that
this is not the most general choice possible, but a relevant and 
convenient one, with obvious extensions.)
We will now look more closely into
the situation where $\lvert S \rvert \nearrow \infty$.
Consider a family of sets 
\begin{equation}\label{eq:S}
   S=S(N), \quad S \subset \ZZ^d, \quad \text{so that} \quad
   \lvert S \rvert \sim N^d \quad \text{as} \;  N \to \infty, 
\end{equation}
where we suppress
once again the dependence of $S$ on $N$.
A reasonable setup is then obtained if
$\frac{1}{N} \cdot S \subset D$, where $D$ is a compact
domain in $\RR^d$, $\frac{1}{N} \cdot S$ becomes dense in $D$ 
for $N \to \infty$,
and there exist functions $\E$ and $f_k$ from  $C^2_b(D,\RR)$ 
(i.e., twice continuously differentiable with bounded second derivatives)
with 
\begin{equation}\label{eq:contapproxe}
   \E_i=\E \Big ( \frac{i}{N} \Big ) + \cO \Big ( \frac{1}{N} \Big )
\end{equation}
and
\begin{equation}\label{eq:contapproxf}
  \F_{ij} =f_{k} \Big ( \frac{i}{N} \Big ) + \cO \Big ( \frac{1}{N} \Big ) \,,
\end{equation}
where $k = j-i$, and the constant in the $\cO(1/N)$ bound
is uniform for all $i$ and $j$. More generally,
one can replace $\cO(1/N)$ in \eref{contapproxe} and 
\eref{contapproxf} by $\cO(1/\eta(N))$ for some function
$\eta(N)$ that grows with $N$, if that better suits the individual
situation.
(Note that our notation is slightly abusive in that $E$ denotes both
the matrix defined by \eref{decomp2}, and the function approximating
its elements; however, the meaning is always obvious from the
context.)

Our main result will be the following theorem. For $S \subset \ZZ^d$,
we will use throughout the shorthand notation
$S-i := \{ j-i \mid j \in S\}$.

\begin{theorem}\label{theo:main} Assume that  $E_i$ and $F_{ij}$
are as in Eqns. \eref{contapproxe} and
\eref{contapproxf} . Assume further that  the 
$C^2_b(D,\RR)$ function $E$ assumes 
its absolute maximum in
${\rm int}(D)$, and that  $f$ satisfies 
\begin{equation} \label{eq:decayfk}
    \sum_{k \in S-i} f_k \Big ( \frac{i}{N} \Big ) 
    \lvert k^{}_{\ell} \rvert k_m^2 \leqslant C
\end{equation}
for some constant $C$,
uniformly for all $i \in S$, and \/ $1 \leqslant \ell,m \leqslant d$.
Then, there exist  constants $0 \leqslant C',C'' <\infty$ such that
\begin{equation} \label{eq:bounds}
 \E(x^*) - \frac{C'}{N} \leqslant \lambdamax \leqslant \E(x^*) + \frac{C''}{N},
\end{equation}
where $x^*$ is a point where $E(x)$ assumes its maximum. 
\end{theorem}

\begin{remark}
It will become clear when we proceed that  the condition on
the derivatives of $E(x)$ and the $f_k(x)$ may be relaxed;
it is indeed sufficient that these functions be continuous and
{\em locally} $C_b^2$, in a neighbourhood of $x^*$.
\end{remark}

Note that the upper bound is clear in view of 
Eqns.~\eref{contapproxe} and \eref{upbound} (recall that
the quadratic form defined by $\F$ is negative semidefinite);
it can be made sharper if the order of the approximation in 
\eref{contapproxe} and \eref{contapproxf} is improved. 
It remains to prove the lower bound (which cannot be improved by
sharpening the $\cO(1/N)$ in \eref{contapproxe} and 
\eref{contapproxf}).
We will do so by evaluating the quadratic form in \eref{RR.1} for
a sequence of test functions of Gaussian type
centred around  $x^*$ in the interior of $D$ 
(and approaching a Dirac measure located
at $x^*$ with increasing $N$).
Specifically, we will use throughout
\begin{equation}\label{eq:testfunc}
   v_i  :=  c e^{-\alpha N \lvert i/N-x^* \rvert^2} \quad \text{with} \;
   c= c(N) \quad \text{s.t.} \quad \sum_{i \in S} v_i^2=1,
\end{equation}
where $\alpha>0$ is a positive real number independent of $N$.

We will first consider the diagonal part and show
\begin{proposition}\label{prop:E}
Let $E_i$ be as in \eref{contapproxe}, and let $x^*$ be a point 
in the interior of $D$ where $E(x)$ assumes its maximum.
Let the $v_i$ be as in Eq.~\eref{testfunc}.
Then,
\[ 
   \sum_{i \in S} E_i v_i^2 = E(x^*) + \cO \Big ( \frac{1}{N} \Big )\,.
\] 
\end{proposition}

The upper bound in the proposition being immediate,  we only need
to prove the lower bound. We will use the following

\begin{lemma}\label{riemann} 
Let $g \! : \, \RR^d \longrightarrow\RR_{\geqslant0}$ be a non-negative,
continuous, integrable function with
$g(x) \leqslant C/(1+ \lvert x \rvert)^{d+\varepsilon}$ for all $x$, and
(fixed) positive constants $C$ and $\varepsilon$.
Then, for any $x^*\in \RR^d$, 
\begin{equation}\label{eq:riem-1}
    \lim_{n\to\infty} \frac{1}{n^d} \sum_{i\in \ZZ^d} g 
    \Big ( \frac{i}{n} - nx^* \Big ) = \int_{\RR^d} g(x) \dd x\,.
\end{equation}
\end{lemma}
\begin{proof} 
Note first that the sum in \eref{riem-1} exists for arbitrary, but
fixed $n$ due to the assumed decay condition for $g$.
Let $b_n := \mbox{\Large $\times$}_{k=1}^d (-1/2n,1/2n]$.
Then, one has $\RR^d = \dot \bigcup_{i\in\ZZ^d} (i/n+b_n)$,
and, for all $x$, there is a (unique) element  $\gamma$ of 
$\ZZ^d/n$ with $x \in (\gamma + b_n)$;  this will be called
$\gamma_n(x)$. 
We now define
\begin{equation} 
   g^+_n(x) : =  \sup_{z \in (\gamma_n(x)+b_n)} g(z), \quad
   g^-_n(x) : =  \inf_{z \in (\gamma_n(x)+b_n)} g(z)\,.
\end{equation}
Since integration over $\RR^d$ is invariant under a shift of argument,
and $g_n^{\pm}$ are step functions, we have
\begin{eqnarray}\nonumber 
    \int_{\RR^d} g^-_n(x) \dd x  
    & = & \int_{\RR^d} g^-_n(x-n x^*) \dd x 
          = \frac{1}{n^d} \sum_{i\in \ZZ^d} g_n^-(i/n-nx^*) \\ 
    & \leqslant &  \frac{1}{n^d} \sum_{i\in \ZZ^d} g(i/n-nx^*) 
     \leqslant  \frac{1}{n^d} \sum_{i\in \ZZ^d} g_n^+(i/n-nx^*) 
     \label{eq:riem-3} \\
   & = & \int_{\RR^d} g^+_n(x-n x^*) \dd x 
         = \int_{\RR^d} g^+_n(x) \dd x\,. \nonumber
\end{eqnarray}
Both $g_n^+$ and $g_n^-$  converge to $g$ pointwise  
(since $g$ is continuous). Furthermore,  
$g_n^{\pm}(x)$ are both bounded from above due to the
properties of the assumed majorizing function, and hence
$\int_{\RR^d} g^-_n(x) \dd x$ and $\int_{\RR^d} g^+_n(x) \dd x$ both
converge to $\int_{\RR^d} g(x) \dd x$ as $n \to \infty$
by the dominated convergence theorem. But then, the
same must be true of the sum in \eref{riem-3}, which proves
the assertion. 
\end{proof}
\begin{corollary}\label{riemann-cor}
For any non-negative integer $k$,
and any $\alpha>0$
\begin{equation}\label{eq:riem-4}
   \lim_{N\to\infty} N^{(k-d)/2}  \sum_{i\in \ZZ^d} 
   \Big \lvert \frac{i}{N} -x^* \Big \rvert^k
   e^{-\alpha N \lvert i/N-x^* \rvert ^2} =
   \int_{\RR^d}  \lvert x \rvert^k e^{-\alpha |x|^2} \dd x\,.
\end{equation}
\end{corollary}

\begin{proof}
Use Lemma \ref{riemann} with $n=\sqrt N$ and 
  $g(x)=|x|^ke^{-\alpha \lvert x \rvert^2}$.
\end{proof}

\begin{lemma}\label{decay} For any 
$A \subset \ZZ^d$, $\delta >0$ and $k\in \NN$,
\begin{equation}\label{eq:decay}
  N^{(k-d)/2} 
  \sum_{\substack{i\in A \\ \lvert i/N-x^* \rvert \geqslant\delta}} 
  \Big \lvert \frac{i}{N}-x^* \Big \rvert^k 
  e^{-2\alpha N \lvert i/N-x^*  \rvert^2}
   = \cO \big(e^{-\alpha N \delta^2}\big)\,.
\end{equation}
\end{lemma}
\begin{proof}
Just note that 
\begin{multline}\label{eq:decay.1}
 N^{(k-d)/2}
  \sum_{\substack{i\in A \\ \lvert i/N -x^*  \rvert \geqslant\delta}} 
  \Big \lvert \frac{i}{N} -x^* \Big \rvert^k 
  e^{-2\alpha N \lvert i/N-x^* \rvert^2}\\
  \leqslant e^{-\alpha N \delta^2}
   N^{(k-d)/2}\sum_{i\in \ZZ^d} \Big | \frac{i}{N}-x^* \Big |^k 
   e^{-\alpha N |i/N-x^*|^2}
\end{multline}
and apply Corollary \ref{riemann-cor} to the last expression to get
the assertion. 
\end{proof}

\begin{corollary}\label{riemann-cor-prime}
  Corollary $\ref{riemann-cor}$ holds true with \/ $\ZZ^d$ replaced by
  $S(N)$ of \eref{S}.
\end{corollary}
\begin{proof}
Since $x^* \in \text{int}(D)$, we may choose a $\delta > 0$ so that
$\ZZ^d \setminus S(N) \subset 
\{i \in \ZZ^d: \lvert i/N - x^* \rvert \geqslant\delta \}$. Then, the
difference in the sum in \eref{riem-4} is $\cO(e^{-\alpha N \delta^2})$,
according to Lemma \ref{decay}, with $A=S(N)$.
\end{proof}

\begin{proof}[of Proposition \ref{prop:E}]
Since we may write
\[
  \Big \lvert \frac{i}{N} - x^* \Big \rvert^k v_i^2 =
  \frac{1}{N^{k/2}} \frac{N^{(k-d)/2} \lvert i/N-x^* \rvert^k 
  e^{-2\alpha N \lvert i/N-x^* \rvert^2}}
  {N^{d/2} \sum_{j\in S}  e^{-2\alpha N \lvert i/N-x^* \rvert^2}}\,,
\]
Lemma \ref{decay} and Corollary \ref{riemann-cor-prime}  
entail that, for $k>0$,
\begin{equation}\label{eq:sumvi-outside}
  \sum_{\substack{i \in S(N) \\ \lvert i/N - x^* \rvert \geqslant\delta}}
  \Big \lvert \frac{i}{N} - x^* \Big \rvert^k v_i^2
  = \cO(e^{-\alpha N \delta^2})
\end{equation}
and
\begin{equation}\label{eq:sumvi-inside}
  \sum_{\substack{i \in S(N) \\ \lvert i/N - x^* \rvert < \delta}}
  \Big \lvert \frac{i}{N} - x^* \Big \rvert^k v_i^2
  = \cO \Big (\frac{1}{N^{k/2}} \Big )\,.
\end{equation}
So far, we have only used that $x^*$ is in $\text{int}(D)$. 
But $x^*$ is also a point where $E(x)$ assumes its maximum, and $E(x)$
is  twice differentiable in a neighbourhood of $x^*$. Hence,
there
exist $\delta>0$ and $0\leqslant C<\infty$, such that, for all $|x-x^*|<\delta$,
$E(x)\geqslant E(x^*)-C |x-x^*|^2$.
Therefore, 
\begin{eqnarray*}
    \sum_{i\in S}v_i^2 E_i 
  & = & \cO \Big (\frac1N \Big ) +
    \sum_{\substack{i \in S \\ \lvert i/N - x^* \rvert < \delta}}
    E \Big ( \frac{i}{N} \Big ) \, v_i^2  
  + \sum_{\substack{i \in S \\ \lvert i/N - x^* \rvert \geqslant\delta}}
    E \Big (\frac{i}{N} \Big ) \, v_i^2  \\
  & \geqslant& E(x^*) \big (1+\cO(e^{-\alpha N \delta^2}) \big )
          - C\sum_{\substack{i \in S \\ \lvert i/N - x^* \rvert < \delta}}
           \Big \lvert \frac{i}{N} - x^* \Big \rvert^2 v_i^2 \\
    &&   + \cO \Big (\frac1N \Big )   + \inf_{x \in D} \big ( E(x) \big ) 
          \sum_{\substack{i \in S \\ \lvert i/N - x^* \rvert \geqslant\delta}}
          v_i^2  \\
  & = & E(x^*) + \cO \Big (\frac1N \Big ),
\end{eqnarray*} 
where we have used \eref{contapproxe} along with normalization in the first, 
\eref{sumvi-outside} in the second, and \eref{sumvi-outside} 
and \eref{sumvi-inside} 
in the last step. This proves the assertion of Proposition \ref{prop:E}. 
\end{proof}

After  dealing with the diagonal part, we are now ready
to embark on the quadratic form. 
\begin{proposition}\label{prop:F}
Let $F_{ij}$ be as in \eref{contapproxf}, and assume that $f$
satisfies   condition \eref{decayfk} of \/ Theorem $\ref{theo:main}$.
Then,
\[ 
   \sum_{i,j \in S} v_i F_{ij} v_j =  \cO \Big (\frac1N \Big )\,.
\] 
\end{proposition}
\begin{proof}
Evaluating  the difference between 
$\lvert i/N - x^* \rvert^2 = \langle i/N - x^*,i/N-x^* \rangle$
and  $\lvert j/N - x^* \rvert^2 = \langle j/N - x^*,j/N-x^* \rangle$,
we first note that
$\lvert j/N - x^* \rvert^2 - \lvert i/N - x^* \rvert^2
= \langle (i+j)/N - 2 x^*, (j-i)/N \rangle$
(here, $\langle . \,, . \rangle$ denotes the scalar product). In view of
$v_i = c e^{-\alpha N \langle i/N - x^*, i/N - x^* \rangle}$, and  with
$j=i+k$,
\[
   v_i > v_{i+k} \, \Longleftrightarrow \,
   \eta(i,k) := \Big \langle \frac{2i+k}{N} - 2 x^*, \frac{k}{N} \Big \rangle 
   > 0
\]
(note that $\eta(i,0)=0$).
Using $F_{ij}=F_{ji}$ (see \eref{Fij}), $(v_i - v_j)^2 = (v_j - v_i)^2$,
and $F_{i,i+k} = f_k(i/N) + \cO(1/N)$ (see \eref{contapproxf}), we
can rewrite the quadratic form as
\begin{eqnarray*}
   \sum_{i,j \in S} v_i F_{ij} v_j 
 & = &  - \frac12 \sum_{i \in S} \sum_{k \in S-i} F_{i,i+k} (v_i - v_{i+k})^2\\
 & = &- \sum_{i \in S} \sum_{\substack{k \in S-i \\ \eta(i,k) > 0}}
                       F_{i,i+k} (v_i - v_{i+k})^2 \\
 & = &  - \sum_{i\in S} \sum_{\substack{k \in S-i \\ \eta(i,k) > 0}}
      \Big ( f_k \Big ( \frac{i}{N} \Big ) 
      + \cO \Big ( \frac{1}{N} \Big ) \Big )
      (v_i - v_{i+k})^2\,.
\end{eqnarray*}
We have thus achieved that the summation includes only terms
where $v_i > v_{i+k}$, which entails that
\[
v_i - v_{i+k}  =  c e^{- \alpha N \lvert i/N - x^* \rvert^2}
                (1 - e^{- \alpha N \eta(i,k)}) \\
               \leqslant  c \alpha N e^{- \alpha N \lvert i/N - x^* \rvert^2}
                        \eta(i,k) \,,
\]
since $1-e^{-x} \leqslant \min(x,1) \leqslant x$ for $x \geqslant0$ (of which we only
use the latter inequality). 
Together with the fact that the quadratic form is negative semidefinite,
this gives
\begin{eqnarray}
  0 & \geqslant& - \frac12 \sum_{i \in S} \sum_{k \in S-i} 
       F_{i,i+k} (v_i - v_{i+k})^2 \nonumber \\
    & \geqslant& - \alpha^2 N^2\sum_{i \in S} v_i^2 
             \sum_{\substack{k \in S-i \\ \eta(i,k) > 0}}
             \Big ( f_k \Big ( \frac{i}{N} \Big )
             + \cO \Big ( \frac{1}{N} \Big ) \Big )  
             \big ( \eta(i,k) \big )^2 \nonumber \\
    & \geqslant& - \alpha^2 N^2\sum_{i \in S} v_i^2 
             \sum_{k \in S-i} 
             \Big ( f_k \Big ( \frac{i}{N} \Big ) 
             + \cO \Big ( \frac{1}{N} \Big ) \Big )  
             \big (\eta(i,k) \big )^2\,. \label{eq:bigsum}
\end{eqnarray}
In the last step, the constraint on the sum could be removed
since we added  to the sum nonnegative terms only: 
$f_k(i/N) \geqslant0$ for $k \neq 0$ (up to $\cO(1/N)$), and
$\big ( \eta(i,k) \big )^2 \geqslant0$ with
equality for $k=0$.

We now note that  \eref{decayfk} entails that, for 
$1 \leqslant \ell,m \leqslant d$,
\begin{equation}\label{eq:fsums}
  \sum_{k \in S-i}  f_k \Big ( \frac{i}{N} \Big ) k_{\ell} k_m, \quad
  \sum_{k \in S-i}  f_k \Big ( \frac{i}{N} \Big ) k_{\ell} k_m^2, \quad 
  \text{and} \quad
  \sum_{k \in S-i}  f_k \Big ( \frac{i}{N} \Big ) k_{\ell}^2 k_m^2/N
\end{equation}
are all bounded from above by
a positive constant $C$ (the latter case relies on $S/N \subset D$ with
compact $D$). Writing
\begin{multline*}
\big ( \eta(i,k) \big )^2 
      =  \Big \langle 2 \Big (\frac{i}{N} - x^* \Big ) + \frac{k}{N}, 
         \frac{k}{N}  \Big \rangle^2 \\
      = \frac{1}{N^2}  \sum_{\ell,m=1}^d k^{}_{\ell} k^{}_m 
      \Big [ 4 \Big (\frac{i^{}_{\ell}}{N} - x_{\ell}^* \Big ) 
      \Big (\frac{i^{}_m}{N} - x_m^* \Big )
        + 4 \Big (\frac{i^{}_{\ell}}{N} - x_{\ell}^* \Big ) \frac{k^{}_m}{N} 
        + \frac{k^{}_m k^{}_{\ell}}{N^2} \Big ] 
\end{multline*}
allows us to bound the various parts
of the sum in \eref{bigsum} as follows:
\begin{multline}\label{eq:term1}
 - 4 \sum_{i \in S} v_i^2 
    \sum_{k \in S-i} f^{}_k \Big ( \frac{i}{N} \Big ) 
    \sum_{\ell,m=1}^{d} k^{}_{\ell} k^{}_m
          \Big ( \frac{i^{}_{\ell}}{N} - x_{\ell}^* \Big ) 
          \Big ( \frac{i^{}_m}{N} - x_m^* \Big )   \\
    \geqslant -4C d  \sum_{m=1}^{d} \sum_{i \in S}
          \Big (\frac{i^{}_{m}}{N} - x^*_{m} \Big )^2 v_i^2 
     =     \cO \Big (\frac{1}{N} \Big )\,, 
\end{multline}
where we used the Cauchy-Schwarz inequality
for 
\[ \sum_{\ell,m=1}^{d}   k^{}_{\ell} k^{}_m
          \Big ( \frac{i^{}_{\ell}}{N} - x_{\ell}^* \Big ) 
          \Big ( \frac{i^{}_m}{N} - x_m^* \Big ) 
          \leqslant \sum_{\ell=1}^d k_\ell^2 \sum_{m=1}^d 
          \Big ( \frac{i^{}_m}{N} - x_m^* \Big )^2,
\]
\eref{fsums}  in the first, and \eref{sumvi-outside}
and \eref{sumvi-inside} in the last step.
 
Again, with \eref{fsums}, \eref{sumvi-outside}, and
\eref{sumvi-inside}, we obtain
\begin{multline}\label{eq:term2}
  - 4 \sum_{i \in S} v_i^2 \sum_{\ell,m=1}^d 
   \sum_{k \in S-i} f_k \Big (\frac{i}{N} \Big ) \frac{k_{\ell} k_m^2}{N} 
   \Big ( \frac{i_{\ell}}{N} - x_{\ell}^* \Big )  \\
   \geqslant  - 4 \frac{C d}{N} \sum_{i \in S} v_i^2 
            \sum_{l=1}^d 
          \Big \lvert \frac{i^{}_{\ell}}{N} - x_{\ell}^* \Big \rvert 
       = \cO \Big ( \frac{1}{N^{3/2}} \Big )\,, 
\end{multline}
where we further used that 
$\sum_{\ell=1}^d \lvert i^{}_{\ell}/N - x_{\ell}^* \rvert
\leqslant  c \ts \lvert i/N - x^* \rvert$ for some positive constant $c$.
Finally, \eref{fsums} also gives that
\begin{equation}\label{eq:term3}
  \sum_{i \in S} v_i^2 \sum_{\ell,m=1}^{d} 
  \sum_{k \in S-i} f^{}_k \Big ( \frac{i}{N} \Big ) 
  \frac{k_{\ell}^2 k_m^2}{N^2} = \cO \Big ( \frac{1}{N} \Big ).
\end{equation}
Combining \eref{term1}, \eref{term2}, and \eref{term3}, we arrive at the
assertion. 
\end{proof}

\begin{remark}
Eq.\ \eref{term3} is the reason that the lower bound in \eref{bounds}
cannot be improved by better approximations in \eref{contapproxe}
and \eref{contapproxf}.
\end{remark}

\begin{remark}
We have, so far, assumed that $x^*$ is in the interior of $D$.
If $x^*$ is on the boundary of $D$, a similar approach may be
taken with a one-sided, exponentially decaying test function. The
error in the approximation will, however,  be 
larger than in the case tackled here.
%In both cases, much finer results can be obtained using more 
%advanced methods of perturbation theory \cite{Kato95},
%which, however, require much more work. 
\end{remark}

So far, we have used the Rayleigh-Ritz variational principle
\eref{RR} to obtain results on the leading eigenvalue of
$H$, but said nothing about the maximizer (note that the latter
need {\em not} coincide with the test function $v$). Recall from Section
\ref{sec:rev} that, for finite $N$,
the maximizer is unique and -- in its $L^1$ version --
given by the ancestral distribution $a=(\h_i \z_i)_{i \in S}$.
Actually, from the bounds above, we can also conclude that $a$
is concentrated in a neighbourhood of $x^*$,
where the width 
of the neighbourhood depends on the behaviour of
$\E$ near its maximum. In the generic case of a quadratic maximum,
$a$ is concentrated in a region with a width of order $1/\sqrt{N}$. 
More precisely, we have:

\begin{theorem}\label{theo:a}
Let $\E_i$ and $\F_{ij}$ satisfy the 
assumptions  of Theorem $\ref{theo:main}$.
Assume  that  $\E$ assumes its maximum at a unique point 
$x^* \in {\rm int}(D)$,
and that the Hessian of $E$ at $x^*$ is 
 negative definite. 

Then, there is a
$\rho > 0$  independent of $N$,  so that, 
for every  $0 < \beta \leqslant 1$ and $N$ large enough:
\[
  \sum_{\substack{i \in S \\ 
                  \lvert i/N - x^* \rvert \geqslant\sqrt{\rho/\beta N}}}
  \!\!\!\!\!\!\!\! a_i \leqslant \beta\,,
\]
where $a$ is the ancestral distribution $($of \eref{a} 
and Prop.~$\ref{prop:a})$. 
\end{theorem}
\begin{proof}
Recall first that the ($L^2$) maximizer of \eref{RR} is given by 
$w = (\sqrt{a_i})_{i \in S}$ (cf.\ \eref{a}). Hence, by 
Theorem \ref{theo:main},  the negative semidefiniteness of $F$, and
\eref{contapproxe}, we have
\begin{equation}\label{eq:maxim}
\begin{split}
  E(x^*) - \frac{C'}{N} 
 & \leqslant \lambdamax = 
   \sum_{i,j \in S} w_i F_{ij} w_j + \sum_{i \in S} E_i w_i^2 \\ 
 & \leqslant \sum_{i \in S} E_i w_i^2 \leqslant \max_{i \in S} E_i
   = E(x^*) + \cO \Big ( \frac{1}{N} \Big ) \,.
\end{split}
\end{equation}
Now, consider $E(x)$ in a neighbourhood of $x^*$. Since the
Hessian at $x^*$ is 
 negative definite, we have 
$E(x) \leqslant E(x^*) - C \lvert x - x^* \rvert^2$ for some $C>0$
in a neighbourhood of $x^*$, this being independent of $N$. For $\varepsilon$
small enough and $\delta(\varepsilon) := \sqrt{\varepsilon/C}$,
therefore, 
\[
  E(x) \leqslant 
  \begin{cases} 
    E(x^*), & \lvert x - x^* \rvert < \delta(\varepsilon) \\
    E(x^*) - \varepsilon, & \lvert x - x^* \rvert
    \geqslant\delta(\varepsilon).
  \end{cases}
\]
Together with \eref{contapproxe} and \eref{maxim}, this implies 
\[
\begin{split}
  E(x^*) + \cO \Big (\frac{1}{N} \Big ) 
 & = \sum_{i \in S} E_i w_i^2 \leqslant E(x^*) 
   - \varepsilon  \!\!\!\!\!\!\!\!
   \sum_{\substack{i\in S \\
         \lvert i/N - x^* \rvert \geqslant \delta(\varepsilon)}}
   \!\!\!\!\!\!\!\!\! w_i^2 + \cO \Big (\frac{1}{N} \Big ) \\
  & \leqslant E(x^*) + \cO \Big (\frac{1}{N} \Big )\,.
\end{split}
\]
Hence, for some positive constant $\gamma$, 
\[
  0 \leqslant \varepsilon 
  \sum_{\substack{i \in S \\  
  \lvert i/N - x^* \rvert \geqslant \sqrt{\varepsilon/C}}}
  \!\!\!\!\!\! w_i^2
  \leqslant \gamma/N
\]
for all sufficiently small $\varepsilon$.
Choosing $\varepsilon = \gamma/\beta N$  and $\rho=\gamma/C$
gives the assertion.
\end{proof}

\begin{remark}
For notational simplicity, we have assumed above that $E(x)$ assumes
its (absolute) maximum at a unique point $x^*$, which is the
generic case. It is obvious from
the proof, however, that an analogous result holds if the maximum
is assumed at a finite number of points (each with a negative
 definite
Hessian). Then, the ancestral
distribution is concentrated on the union of the
corresponding neighbourhoods of  these points (or a subset thereof),
again with widths of order $1/\sqrt{N}$. 
\end{remark}

Let us return to the case where $E(x)$ assumes
its (absolute) maximum at a unique point $x^*$.
We have seen that the  ancestral distribution
concentrates around $x^*$ for $N \to \infty$, in the sense that
any given fixed fraction $1-\beta$ (or even more) of the 
distribution's mass is contained in
a region whose width decreases with $1/\sqrt{N}$.
From this, we can further conclude that 
the {\em mean ancestral type} (in proper scaling), 
$(\sum_i i a_i)/N$, converges to $x^*$, which adds some interpretation
to the maximum principle in Theorem \ref{theo:main}.
More precisely, we have 

\begin{corollary}\label{cor:mean}
Under the assumptions of Theorem $\ref{theo:a}$, we have
\[
   \sum_{i \in S} \frac{i}{N} a_i = x^* + 
   \cO \Big ( \frac{1}{N^{1/3}} \Big )\,,
\]
as $N \to \infty$.
\end{corollary}
\begin{proof}
By the triangle inequality, and with a constant $\rho$ as in Theorem
\ref{theo:a}, we have
\begin{eqnarray*}
 \Big \lvert \sum_{i \in S} \frac{i}{N} a_i - x^* \Big \rvert & = &
 \Big \lvert \sum_{i \in S} \Big ( \frac{i}{N} - x^* \Big ) a_i \Big \rvert 
  \leqslant \sum_{i \in S} \Big \lvert \frac{i}{N} - x^* \Big \rvert a_i \\
  & = & 
  \sum_{\substack{i \in S \\ \lvert x^* - i/N \rvert < \sqrt{\rho/\beta N}}} 
  \!\! \Big \lvert \frac{i}{N} - x^* \Big \rvert a_i 
    + \sum_{\substack{i \in S \\ 
  \lvert x^* - i/N \rvert \geqslant \sqrt{\rho/\beta N}}} 
  \!\! \Big \lvert \frac{i}{N} - x^* \Big \rvert a_i
\end{eqnarray*}
for all $0 < \beta \leqslant 1$. 
The first term is bounded by $\sqrt{\rho/\beta N}$ by construction.
Due to Theorem \ref{theo:a} and the fact that $S/N \subset D$ with
compact $D$, the second term is bounded by $C \beta$ for some positive 
constant $C$. Thus,
\[
 \Big \lvert \sum_{i \in S} \frac{i}{N} a_i - x^* \Big \rvert
 \leqslant \sqrt{\frac{\rho}{\beta N}} + C \beta
\]
for all $0 < \beta \leqslant 1$ and $N$ large enough. Choosing
$\beta = \beta(N) = 1/N^{1/3}$ gives the assertion.
\end{proof}

\begin{remark}
So far, we have only considered the leading eigenvalue and the
corresponding eigenvector, in `crudest' approximation order
$1/N$. Using more advanced techniques from perturbation theory
\cite{Kato95}, it would be possible to obtain results on further
eigenvalues and eigenvectors, as well as higher-order error terms.
\end{remark}

\section{Lumping}
\label{sec:lump}

Let us now drop the specific assumptions of the previous
section, 
return to the general situation in the Introduction,
and reflect on the type space $S$,
which has, so far, remained unspecified.
In the example of Section \ref{sec:ex}, the types were
defined in terms of some intermediate genetic level that
could be derived from a more detailed picture.
In this Section, we will show that a large class of  models 
on some type space $S$ 
can be derived, in a natural  way, from 
models defined on a `larger' space $\fS$ 
(to be called genotype space) if  the branching and mutation rates
satisfy certain symmetry or compatibility conditions.
The idea rests on the common assumption that fitness depends on the
genotype through an intermediate level of `effective' parameters
(which may, for example, be `phenotypes', or
`genetic values' in quantitative genetics), and
the mapping from the genotype to this intermediate level is multiple-to-one.
One will therefore try and combine several of the genotypes
into a single effective type; if this is 
also compatible with the mutation scheme, a
reduction of the number of dimensions is possible. In the
theory of Markov chains, this approach is known as {\em lumping}
\cite[Ch.\ VI]{KeSn81}.
We will proceed in two steps: First, the lumping procedure will be
described in an abstract setting, with arbitrary genotype and
type spaces $\fS$ and $S$, respectively. In a second step,
we will specialize to the concrete sequence (or multi-locus)
picture.

For the first step, let $\fS$ be a possibly large, but finite set.
In analogy with \eref{lindgl}, consider the dynamics 
\begin{equation}\label{eq:lindgl2}
  \dot \rho =\rho (\cM + \cR)
\end{equation}
on $\RR^{|\fS|}$, with $\cM$ a Markov generator and 
$\cR=\text{diag}\{\cR_{\sigma} \; | \; \sigma \in \fS\}$.
For this discussion, $\cM$ need neither be irreducible nor reversible.

\begin{figure}
  \centerline{\input{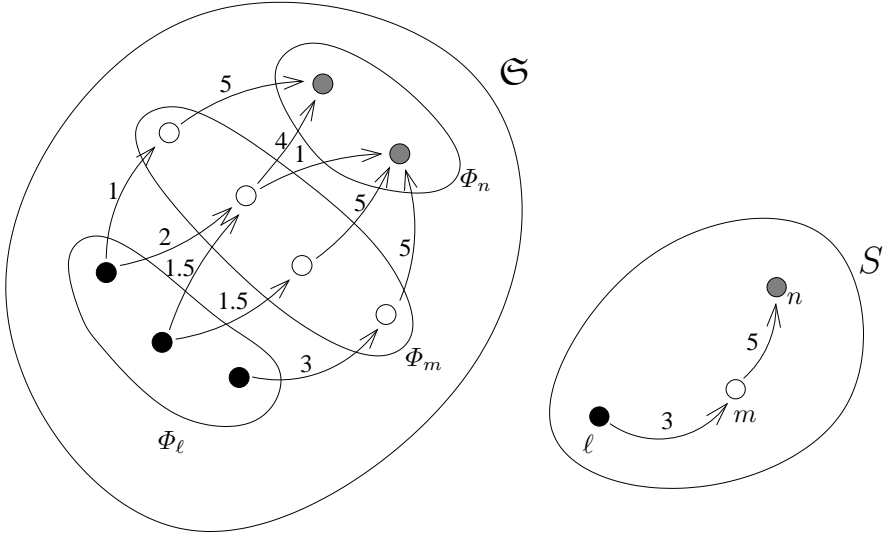}}
  \label{fig:lumping}
  \caption{The lumping procedure. The `large' space $\fS$ is partitioned
so that all elements in a given subset, say $\Phi_m$, have the same
reproduction rate $R_m$ (Eq.\ \eref{rlump}), and the same total
mutation rate, $\sum_{\tau \in \Phi_n} \cM_{\sigma,\tau}$,
to elements in any other given subset $\Phi_n$
(Eq.\ \eref{mlump}). Then, each subset may be represented by a single
element in a smaller space $S$, and the induced `effective' model on
$S$ is again a linear mutation-reproduction model.}
\end{figure}

Consider a mapping
\begin{equation}\label{eq:phi}
   \varphi: \; \fS \longrightarrow S=\text{im}(\varphi)\,
\end{equation}
so that $\fS$ may be understood as the disjoint union of  fibres $\Phi_m$:
\[
  \fS = \dot \bigcup_{m \in S} \; \Phi_m\,, \quad \text{with} \quad
  \Phi_m := \{ \sigma \in \fS \; | \; \varphi(\sigma)=m \} = \varphi^{-1}(m)\,.
\]
We will now give conditions under which the dynamics \eref{lindgl2}
may be reduced to a dynamics on $S$. The following result is a
variant of a theorem by Burke and Rosenblatt \cite{BuRo58},
see also \cite[Chapter VI]{KeSn81}. The setting is illustrated  in Figure 
\ref{fig:lumping}. 

\begin{theorem} 
Let $\fS$ and $S$  be finite, let $\varphi$ be the mapping of
\eref{phi}, and assume that
there are matrices $\M=(\M_{nm})_{n,m \in S}$  and 
$R={\rm diag}\{R_i \; | \; i \in S\}$  with
\begin{eqnarray}\label{eq:rlump}
  \cR_{\sigma} & = & R_{\varphi(\sigma)}, \qquad 
                     \text{for all} \; \sigma \in \fS, \\[1mm]
  \label{eq:mlump}
  \sum_{\tau \in \Phi_m} \cM_{\sigma,\tau} & = & \M_{\varphi(\sigma),m}, \quad
                     \text{for all} \; \sigma \in \fS, \; m \in S\,,
\end{eqnarray}
where $\cM$ is the  
Markov generator of Eq.~\eref{lindgl2}.
Then, $\M$ is a Markov generator on $\RR^{\lvert S \rvert}$. If
$\rho$ solves \eref{lindgl2}, then
\begin{equation}\label{eq:yrho}
  \y_m := \sum_{\sigma \in \Phi_m} \rho_{\sigma}\,
\end{equation}
satisfies the differential equation \eref{lindgl}, i.e.,
$ \dot \y_m = \sum_n \y_n (\M_{nm} + R_n \delta_{nm})$.
If $\cM$ 
has stationary distribution $\tpi = (\tpi_{\sigma})_{\sigma\in \fS}$, 
$M$ has stationary distribution $\bpi=(\pi_m)_{m \in S}$, where
$\pi_m = \sum_{\sigma\in\Phi_m} \tpi_{\sigma}$;
reversibility of $\cM$ with respect to $\tpi$ implies that of
$M$ with respect to $\pi$.
If $\cM + \cR$ has principal left eigenvector ${\tilde{h}}$,
$\M+R$ has principal left eigenvector $\h$ with 
$\h_m = \sum_{\sigma\in\Phi_m} {\tilde{h}}_{\sigma}$.
\end{theorem}

\begin{proof}
The proof is a straightforward verification. 
Note first that
$\M$ is a Markov generator (on $\RR^{|S|}$), because, for any 
$\sigma\in \Phi_m$, 
\[
   \sum_{n\in S} \M_{mn}  =  \sum_{n\in S} \sum_{\tau \in \Phi_n} 
   \cM_{\sigma\tau}
    =  \sum_{\tau \in \fS} \cM_{\sigma\tau} \; = \; 0\,,
\]
since $\cM$ is a Markov generator.

Starting now from
\eref{yrho} and \eref{lindgl2}, we find
\begin{eqnarray*}
   \dot \y_m & = & \sum_{\sigma \in \Phi_m} \dot \rho_{\sigma} 
               = \sum_{\sigma \in \Phi_m} \sum_{\tau\in \fS} \rho_{\tau}
                 (\cM_{\tau\sigma} + \cR_{\tau} \delta_{\tau\sigma}) \\
%             & = & \sum_{\tau\in S} \rho_{\tau} 
%                 \Big( \sum_{\sigma\in \Phi_m} \cM_{\tau\sigma}
%                      + R_{\phi(\tau)} \delta_{\varphi(\tau),m} \Big) \\
             & = & \sum_{n\in S} \, \sum_{\tau\in \Phi_n} \rho_{\tau} 
                  \big( \M_{\varphi(\tau),m} + 
                  R_{\varphi(\tau)} \delta_{\varphi(\tau),m} \big) \\
             & = & \sum_{n\in S} \y_n (\M_{nm} + R_n \delta_{nm}) \,,
\end{eqnarray*}
where we have used \eref{rlump} and \eref{mlump} in the 
second step, and \eref{yrho} in the last, together
with the fact that both $\M_{\varphi(\tau),m}$ and 
$R_{\varphi(\tau)} \delta_{\varphi(\tau),m}$ are constant on every fibre
$\Phi_n$.

Finally, the assertions on stationary
distributions and reversibility are direct verifications in the
same spirit.
\end{proof}

\section{From sequence space to type space}
\label{sec:seqspace}

In this Section, we will be more explicit and start from sequence
space. The natural scheme that will emerge  involves the
grouping of  sites together with a `coarse-grained' 
dependence on some `genetic distance'.  
Many of the  frequently-used models fall into this
scheme. Related results appear in statistical physics, compare 
\cite{BEGK01,BABG03}, from where we will borrow some techniques. 

Let us begin with the general setup for a mutation-reproduction 
model on sequence 
space. We will assume that the type $\sigma$ of an individual is 
characterized by 
a (DNA, RNA) sequence which we take to be an element of the space
$\fS := \Sigma^N$ with $\Sigma = \{1,\dots,q\}$; we write
$\sigma = (\sigma_1,\dots,\sigma_N)$. 
For generality, we let $q$ be an 
integer $\geqslant 2$; if $q=2$, the alternative choice
$\Sigma = \{-1,1\}$ is often more convenient. 
Consider now a partition of the  
 set of sites $\Lambda=\{1,\dots,N\}$ into 
$K$ disjoint subsets $\Lambda_k$, i.e., 
\begin{equation}\label{eq:partition}
  \Lambda = \dot \bigcup_{1\leqslant k \leqslant K} \; \Lambda_k.
\end{equation}
Let 
$\cP(\Sigma) = \{ (\mu_1,\dots,\mu_q) \; | \; \mu_{\ell} \geqslant 0, 
\sum_{\ell} \mu_{\ell} = 1 \}$ denote the simplex of
probability measures (or vectors) on $\Sigma$. Set, with obvious meaning,

\[
  \cP_{\Lambda_k}(\Sigma) := \cP(\Sigma) \cap 
  \Big \{ 0, \frac{1}{|\Lambda_k|}, \frac{2}{|\Lambda_k|}, \dots,
             1 - \frac{1}{|\Lambda_k|},1 \Big \}^q
\]
and 
\begin{equation}\label{eq:parspace}
  \parspace(\Sigma) = 
  \bigotimes_{k=1}^K \cP_{\Lambda_k}(\Sigma)\,.
\end{equation}
That is, $\parspace(\Sigma)$ is the set of  product 
measures with 
values restricted to certain rationals induced by the partition.

Consider now the mapping (which will take the role of $\varphi$ from
the previous section) 
\begin{equation}\label{eq:defmag}
  \fm: \; \Sigma^N  \longrightarrow  \QQ^{Kq}, 
       \quad  \sigma    \mapsto  \fm(\sigma) 
\end{equation}
with 
$
  \fm(\sigma) = 
  \big(\fm_k^{\ell}(\sigma)\big)^{1\leqslant\ell\leqslant q}
                                _{1\leqslant k \leqslant K }
$
and 
\begin{equation}\label{eq:mil}
  \fm_k^{\ell}(\sigma) := 
  \frac{1}{|\Lambda_k|} \sum_{s \in \Lambda_k} \delta_{\ell,\sigma_s}
  = \frac{1}{|\Lambda_k|} \, 
    \big | \{ s \mid  s \in \Lambda_k,\sigma_s=\ell\} \big |\,.
\end{equation}

So, $\fm_k^{\ell}(\sigma)$ is the fraction of the sites 
in $\Lambda_k$ which are in state $\ell \in \Sigma$.
Note that  
$\sum_{\ell=1}^q \fm_k^{\ell}(\sigma)=1$, i.e., for each $k$, 
$\fm_k(\sigma):= \big(\fm_k^1(\sigma),\dots,\fm_k^q(\sigma)\big)$ 
defines a probability measure on $\Sigma$, with 
$\fm_k \in \cP_{\Lambda_k} (\Sigma)$.

Describing the system in terms of these lumped quantities
will only lead to a simplification if a suitable symmetry is available.
In our case, this is given by those permutations of the sites
that are compatible with the chosen partition.

Let $\G_{\Lambda}$ be the permutation group on $\Lambda=\{1,\dots,N\}$,
i.e.,
\[
  \G_{\Lambda} := \{ \gamma \; | \; 
  \gamma:\; \Lambda \rightarrow \Lambda \; \text{is a bijection} \}\,,
\]
and $\G_{(\Lambda_1,\dots,\Lambda_K)}$ the subgroup compatible with
the partition \eref{partition}, i.e., 
\[
 \G_{(\Lambda_1,\dots,\Lambda_K)} = 
 \big \{ \gamma \in \G^{}_{\Lambda} \mid 
 \gamma(\Lambda_k) = \Lambda_k, 1 \leqslant k \leqslant K \big \} 
 \simeq \G_{\Lambda_1} \times \dots \times \G_{\Lambda_K}\,.
\]
We introduce the canonical action of the permutation group 
on $\Sigma^N$ through the
inverse permutation of sites, i.e.,
$(\gamma \sigma)_j = \sigma_{\gamma^{-1}(j)}$. We are now ready for
\begin{theorem}\label{theo:seqlump}
Let $\Sigma^N=\{1,\dots,q\}^N$, and matrices 
$\cM=(\cM_{\sigma,\tau})_{\sigma,\tau\in\Sigma^N}$ and
$\cR={\rm diag}\{\cR_{\sigma} \mid \sigma \in \Sigma^N\}$ be given,
with $\cM$ a Markov generator. 
Let $\rho$ solve $\dot \rho = \rho (\cM+\cR)$.
Furthermore, let \/ $\fm$ be as in \eref{defmag},
and 
$\hat S := \fm(\Sigma^N) \subset \QQ^{Kq}$. 
Assume  that there exist a function
$g \! : \, \Sigma^N\times\Sigma^N \longrightarrow\RR_{\geqslant 0}$,
and matrices $\hM=(\hM_{mn})_{m,n\in \hat S}$ and 
$R={\rm diag}\{R_n \;| \; n \in \hat S\}$, so that the following conditions
are satisfied:
\begin{itemize}
\item[\rm(a)]
$\;\f(\gamma\tau,\gamma\sigma) = \f(\tau,\sigma), \quad \text{for all} \;
                          \gamma \in \G_{(\Lambda_1,\dots,\Lambda_K)}\,;$
\item[\rm(b)]
$\; \mutseq_{\sigma \tau}=\hM_{\fm(\sigma),\fm(\tau)} \ts g(\sigma,\tau), \quad
\text{for all} \; \sigma, \tau \in \Sigma^N\,;$
\item[\rm(c)]
$\; \braseq_{\sigma}=R_{\fm(\sigma)}, \quad \text{for all} \; 
\sigma \in \Sigma^N$. 
\end{itemize}
Then,
$\y_m := \sum_{\sigma \in \Phi_m} \rho_{\sigma}$ solves the differential
equation $\dot y = y (M+R)$, where
$$
\M_{nm} = \hM_{nm} \sum_{\tau\in \Phi_m} \f(\sigma,\tau)
$$
is independent of the choice of $\sigma \in \Phi_n$. Moreover,
$\stoma$ is a Markov generator. 
If $\cM$ 
has stationary distribution $\tpi = (\tpi_{\sigma})_{\sigma\in \fS}$, 
$M$ has stationary distribution $\bpi=(\pi_m)_{m \in S}$, where
$\pi_m = \sum_{\sigma\in\Phi_m} \tpi_{\sigma}$;
reversibility of $\cM$ with respect to $\tpi$ implies that of
$M$ with respect to $\pi$.
If \/ $\cM + \cR$ has principal left eigenvector \/ 
${\tilde{h}}=(\tilde{h}_{\sigma})_{\sigma \in \fS}$,
then $\M+R$ has stationary distribution 
$\h=(h_m)_{m\in \hat S}$ with 
$\h_m = \sum_{\sigma\in\Phi_m} {\tilde{h}}_{\sigma}$.
\end{theorem}
\begin{proof}
For 
$\gamma \in \G_{(\Lambda_1,\dots,\Lambda_K)}$, we have
\begin{equation}\label{eq:2cond}
   \fm(\gamma \sigma) = \fm(\sigma) \quad \text{and} \quad
   \gamma(\Sigma^N) = \Sigma^N\,,
\end{equation}
where the first identity is obvious from \eref{mil}.
\Eref{2cond} entails that 
\begin{equation}\label{eq:transit}
  \gamma (\Phi_m) = \Phi_m,
\end{equation} i.e., 
$\G_{(\Lambda_1,\dots,\Lambda_K)}$ acts transitively on $\Phi_m$.

In order to apply Theorem \ref{theo:main}, we have to check 
assumption \eref{mlump}.
Consider therefore 
$\sum_{\tau\in \Phi_m} \cM_{\sigma\tau} = 
\hM_{\fm(\sigma),m} \sum_{\tau \in \Phi_m} \f(\sigma,\tau)$.
For arbitrary 
$\gamma \in \G_{(\Lambda_1,\dots,\Lambda_K)}$,
assumption (a)  and Eq.\ \eref{transit} give
\begin{eqnarray*}
 \psi(\sigma)& := & \sum_{\tau\in\Phim} \f(\sigma,\tau)  =   
  \sum_{\tau\in\Phim} \f(\gamma\sigma,\gamma\tau) \\
   & = &  \sum_{\tau'\in\gamma(\Phim)} \f(\gamma\sigma,\tau') 
     =  \sum_{\tau'\in\Phim} \f(\gamma\sigma,\tau') = \psi(\gamma \sigma)\,. 
\end{eqnarray*}

Due to the transitivity of $\G_{(\Lambda_1,\dots,\Lambda_K)}$ on
$\Phim$, $\psi(\sigma)$ is constant on the fibres 
$\Phi_{\fm(\sigma)}$.
Assumption \eref{mlump} is therefore valid, and an
application of Theorem \ref{theo:main} then gives the desired result.
\end{proof}
\begin{remark} 
The connection with the situation in \sref{maxeval} is  made by 
setting $d=Kq$, and observing that 
$\tilde S/N \subset [0,1]^d =: \tilde D$. Obviously, $\tilde S$ and $\tilde D$
must take the roles of $S$ and $D$. If 
$\lvert \Lambda_k \rvert \sim \alpha_k N$
with positive constants $\alpha_k$, $1 \leq k \leq K$, and
$\sum_k \alpha_k = 1$, then $\tilde S$ becomes dense in $\tilde D$ as
$N \to \infty$. The corresponding $D$ is a parallelepiped with edge lengths
$\alpha_k$.
\end{remark}

Examples of particular relevance emerge if $\f$ is a 
$\G_{(\Lambda_1,\dots,\Lambda_K)}$-invariant
{\em distance}, such as the Hamming distance (i.e., the number of sites
at which two sequences differ). A very simple case was implicit in 
\sref{ex}, where the single-step mutation model 
on $S=\{0,1,\dots,N\}$ was interpreted in terms of a model on
$\{-1,1\}^N$. Here, a site in state $+1$ or $-1$ corresponds to  
a site whose state does or does not coincide with the respective 
state in a reference sequence (sometimes called the `wildtype').
If the reproduction and mutation rates 
only depend on  the
Hamming distance from the reference sequence, we are in a setting with
$K=1$, $q=2$ and hence $d=2$, which further boils down to $d=1$
if the restriction $\fm_1^1+\fm_1^2=1$ is used (see also below). 
In such a simple case, the lumped model is immediate.
More elaborate examples will be discussed in the next Section.

\section{\bf Towards Applications}
\label{sec:applic} 

In many  examples of sequence space models,
the lumping construction as
described in the previous sections leads to an effective model
to which the maximum principle of  \sref{maxeval}
may then be applied. In particular, a given example will be a case
for Theorem \ref{theo:main} if it has the following properties:

\begin{itemize}
\item[(P1)] 

The partition $\{\Lambda_k\}_{k=1}^K$ in \eref{partition}
is relatively uniform, in the sense 
that there exist constants $0<c\leqslant C<1$ such that

\[ 
  c \leqslant \inf_{1\leqslant k \leqslant K}\frac {|\Lambda_k|}N
    \leqslant \sup_{1\leqslant k \leqslant K}\frac {|\Lambda_k|}N \leqslant C
\]
uniformly in $N$. (Alternatively, this may be replaced by the 
single, and slightly weaker, condition  
$\liminf_{N\to\infty}  \inf_{1\leqslant k \leqslant K}
\frac {\lvert \Lambda_k \rvert}N  > 0$; 
note that $\sum_k \lvert \Lambda_k \rvert = N$ by construction.)
This condition ensures that $x_i=i/N$ will become a meaningful continuous 
type variable for $N \to \infty$.  
\end{itemize}
For the next two properties, a suitable enumeration of the
elements of $S$ is required to ensure an appropriate representation
of the matrices $\bM$ and $\bR$.
\begin{itemize} 
\item[(P2)] The function $\f$ that occurs in the sequence space
mutation matrix and that is required in the lumping procedure
(see Theorem \ref{theo:seqlump}) decreases sufficiently fast
away from the diagonal. Note that 
under condition (P1), for any $\sigma,\tau$ we have that
\[
   d_{\rm H}(\sigma,\tau)\geqslant 
   \frac{N}{C} \, \|\fm(\sigma)-\fm(\tau)\|_1\,,
\]
where
$d_{\rm H}$ is the Hamming distance. Thus, if $g$ has compact support
independent of $N$ (as in the example in \sref{ex}),
or if it decays sufficiently fast (e.g., exponentially) with $d_{\rm H}$, 
this entails the decay condition on $f$ in Theorem \ref{theo:main}.

\item[(P3)] After lumping, the effective reproduction and
mutation matrices $R$ and $M$  lend themselves to a 
 continuous 
approximation. That is, 
$R_m = r(m) + \cO(1/N)$ 
and $M_{mn}= s \big (m, n \big ) + \cO(1/N)$  
with  functions $r$ and $s$ that are $C_2^b(D,\RR)$,
where the implied constant in the $\cO(1/N)$ bound
is uniform for all $m$ and $n$. This
entails the approximation condition on $E$ and $F$ in 
\eref{contapproxe} and \eref{contapproxf} that is also required for
Theorem \ref{theo:main}.
\end{itemize}

Clearly, (P2) and (P3) stipulate that the enumeration of the types
is adapted to the problem. The right choice is often intuitively
clear, as in the examples in  \sref{ex}, and in \cite{GaGr04}.
But sometimes more thought
is required, as will be illustrated  by means of a few examples and special
cases below.

\begin{itemize}
\item[(E1)] Some simplifications arise in the case $q=2$, where we
now use $\Sigma=\{-1,1\}$ rather than $\{1,2\}$. Here,
the constraint 
$\fm_k^1+\fm_k^2=1$ can be used to reduce the number of variables per
subset to one. It is  convenient to 
set 
$\magn_k\equiv \macpar_k^1-\macpar_k^2$. Eq.\ \eref{parspace} is then 
replaced by 
\[
\parspace(\Sigma) =  \bigotimes_{k=1}^K\{-1,-1+\frac{2}{|\Lambda_k|},\dots,
                                 1-\frac{2}{|\Lambda_k|},1\}\,,
\]
and we obtain the simple formula 
\[ 
   \magn_k(\sigma)=\frac 1{|\Lambda_k|}\sum_{s\in\Lambda_k}\sigma_s\,.
\]

\item[(E2)] The case $d=1$ (and hence $S \subset \ZZ$)
corresponds to so-called `mean field models'.
They have been studied in the case 
where $\f(\sigma,\tau) = 0$ for $d_{\rm H}(\sigma,\tau) \geqslant 1$,
i.e., mutation is restricted to neighbours in sequence space
(see \cite{BBW97,BBW98,WBG98,BaWa01,HRWB02}  for $q=2$, and 
\cite{HWB01,GaGr04} for  $q=4$). 

\item[(E3)] A special type of models that falls into the above 
class is related to fitness landscapes based on Hopfield Hamiltonians.
These are special cases of spin-glass models \cite{MPV87} that were 
originally motivated by neural networks, 
then became prototype models for random interactions in
statistical physics, and were later also used as tunably rugged
fitness landscapes in biology \cite{Leut87a,Tara92}. 

Let us consider the case $q=2$, with
sequence space $\fS=\Sigma^N=\{-1,1\}^N$. A Hopfield Hamiltonian
then is a function that assigns to every $\sigma \in \fS$
an energy $\cH_N(\sigma)$ in the following way:
$M$  elements $\xi^1,\dots,\xi^M$ of $\Sigma^N$ (known as
{\em patterns}) are specified
(usually by independent random draws from $\Sigma^N$).  Given these, one
defines  
\begin{equation}\label{eq:HN}
  \cH_N(\sigma) := \frac{1}{N} \sum_{\mu=1}^M 
                     \sum_{s,t=1}^N \sigma_s^{} \sigma_t^{} \xi^\mu_s\xi^\mu_t
                     =N \sum_{\mu=1}^M \big ( \ov_\mu(\sigma)\big )^2,
\end{equation}
where 
\begin{equation}\label{eq:omega}
  \ov_{\mu}(\sigma) := \frac{1}{N} \sum_{s=1}^N \sigma_s^{} \xi_s^{\mu}
                     = \frac{1}{N} \ts \langle \sigma, \xi^{\mu} \rangle \,,
\end{equation}

i.e., a sequence  is assigned an energy by sitewise comparison
of the sequence with all patterns (see Fig.\ 2).
The properties (in particular, the ruggedness) of the energy landscape
so defined (and to be used to assign fitness, see below)
depends on the number and the particular choice of
the patterns. 

\begin{figure}
  \centerline{\input{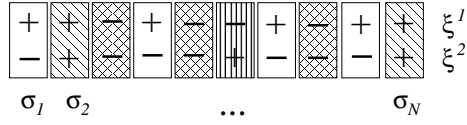}}
  \label{fig:hopfield}
  \caption{
  Lumping in a Hopfield model with $M=2$. Here,
  $\xi^1,\xi^2 \in \{-1,1\}^N$
  are two reference sequences (`patterns'). Fitness is assigned to
  a sequence $\sigma = (\sigma_1,\sigma_2, \ldots, \sigma_N) \in
  \{-1,1\}^N$ by sitewise comparison with both patterns (Eqs.~\eref{HN},
  \eref{omega}, and \eref{Rsigma}).
  This defines four subsets of sites (indicated by different shadings) so
  that the sites in each subset are equivalent with respect to both
  $\xi^1$ and $\xi^2$ and may thus be permuted without a change of
  fitness. 
}
\end{figure}

Let us now explain the lumping procedure for $\cH_N(\sigma)$,
as adopted from \cite{BABG03} and illustrated in Fig.\ 2  
(the more general setting with $q>2$ can be found in \cite{Gayr92}).
To this end, we associate with the collection of row vectors
$\xi^\mu$ the $M\times N$ matrix 
$\xi=(\xi^{\mu}_s)^{1\leqslant\mu\leqslant M}_{1\leqslant s \leqslant N}$. 
We denote by $\xi^\mu$ the 
rows and by $\xi_s$ the columns of this matrix. A partition 
$\Lambda_1,\dots,\Lambda_K$
with $K\leq 2^M$ is now obtained as follows. Let $e_1,\dots, e_{2^M}$ 
$\big ( e_k^{}=(e_k^{\mu})^{1\leqslant\mu\leqslant M} \big )$ denote an
enumeration of all $M$-dimensional column vectors with entries $\pm 1$.
Then we set 
\[ 
   \Lambda_k :=  \{s \in \Lambda \mid  \xi_s=e_k\}\,.
\]
If all the $\Lambda_k$ are non-empty, $K=2^M$; otherwise, empty
subsets may be omitted, and $K < 2^M$. 
We then have 
\[ 
  \ov_\mu(\sigma) = \frac{1}{N} \sum_{k=1}^K e^\mu_k 
                                \sum_{s\in \Lambda_k} \sigma_s^{}
                 = \frac{1}{N} \sum_{k=1}^K  \lvert \Lambda_k^{} \rvert 
                                             e^\mu_k \magn_k^{}(\sigma)\,,
\]
and so 
\[
  \cH_N(\sigma)
  = N\sum_{\mu=1}^M \sum_{k,\ell=1}^K e^\mu_k e^\mu_{\ell} 
    |\Lambda_k^{}||\Lambda_{\ell}^{}|
    \magn_k^{}(\sigma)\magn_{\ell}^{}(\sigma)
\]
is a function of the $\magn_k(\sigma)$. Thus, if we consider 
reproduction and mutation rates of the form 
\begin{eqnarray}
  \mutseq_{\sigma\tau} & = & 
  \alpha \big ( \cH_N(\sigma),\cH_N(\tau) \big ) \,
  \f(\sigma,\tau)\,,  \\ \label{eq:Rsigma}
  \braseq_{\sigma} & = & \beta \big (\cH_N (\sigma) \big)\,,
\end{eqnarray}
with a nonnegative function $\alpha$ and any real function $\beta$, 
we may apply Theorem~\ref{theo:seqlump} to derive the effective 
dynamics with lumping according to the values of $\magn_k(\sigma)$. 
In particular, the
choice $\beta(x)=x$  gives the familiar Hopfield fitness landscape,
and $\alpha(x) \equiv 1$ along with
$\f(\sigma,\tau) = \mu$ for $d_H(\sigma,\tau)=1$, 
$\f(\sigma,\tau) = 0$ for $d_H(\sigma,\tau)>1$, and
$\f(\sigma,\sigma) = - 2 N \mu$
yields  the  decoupled sequence space mutation model where every
site mutates independently and at the same rate $\mu$ (e.g.,
\cite{BaWa01}). It may be considered as the decoupled variant
of the quasispecies model \cite{EMcCS89}; the latter  may be constructed
in a similar way. Both are mutation-selection models in
a molecular setting and  well known for their error thresholds
that may occur when $\mu$ surpasses a critical value. A preliminary
analysis 
of sequence space mutation-selection models with Hopfield fitness
has been given in \cite{Leut87a,Tara92} and shows a
rich behaviour, with various error thresholds, depending on the
specific choice of patterns.

\end{itemize} 

\section{Summary and Discussion}

The motivation for this work came from haploid mutation-selection
models, or other essentially linear models, which frequently
appear in population biology. These are models
for relative frequencies of types (genotypes, age classes...)
in a population, which turn linear after a suitable transformation
to quantities that may be interpreted as the absolute
frequencies that would be obtained if growth were unrestricted.

We have been mainly concerned with the leading eigenvalue of
the matrix that describes this {\em linear} dynamics. This leading
eigenvalue is the key to the asymptotic properties of
the corresponding {\em essentially linear} model.
For example, it directly yields the mutation load in a
mutation-selection model. It also provides the key to the stationary
distribution of types in the present  as well
as the ancestral population (the latter is obtained when running
the process backward into the past until stationarity is reached).
Furthermore,  its parameter dependence  determines whether
error thresholds occur in a given system. 

We have considered here  the large class of models with
a {\em reversible} mutation part, meaning that, in the
(hypothetic) mutation equilibrium $\pi$ in the absence of reproduction,
the mean number of transitions between any
pair of types is the same in the forward and the backward direction.
This is a standard assumption
in many models of population genetics.
Note that any {\em symmetric} mutation generator is automatically
reversible (because $\pi$ is then the equidistribution).
Many mutation models of classical population genetics are reversible
(like  the random-walk mutation model with Gaussian mutant
distribution \cite{Buer00,Redn04}),
and the same holds for practically all models of nucleotide
evolution, as discussed already in Section \ref{sec:intro}.
At the molecular level, reversibility is a basic assumption
on which practically all model-based phylogenetic tree estimation 
methods rest. 
%It
%has been widely tested statistically, and rarely have deviations
%been found \cite{RICE?}. \marginpar{Arndt nach Ref fragen}

Reversibility implies that the matrix $H$
that governs the linear(ized) dynamics is similar to
a symmetric one, which in turn means that its leading
eigenvalue may be determined by the Raleigh-Ritz
variational principle. But this alone is not very
useful in practice if the number of types is large, which
is the usual situation in all but a few textbook examples.
The main concern of this paper, therefore, was to reduce the number
of dimensions to its `effective' number. This involved two steps:
A lumping procedure that leads to an
equivalent smaller, still discrete, system; and an
approximation that turns the discrete system into
a continuous one by replacing the discrete types by
a continuous type variable. Let us discuss them in turn.

\smallskip

{\em Lumping:} This a kind of coarse-graining
that applies if the fitness function and the mutation
model on the 'original' (genotype) space $\fS$ have enough
symmetries to allow for lumping of several states of
$\fS$ into a single one, so that the induced  `effective'
model on a smaller space $S$ is again a mutation-reproduction model.  
As illustrated in Fig.~1, this works if
\begin{enumerate}
\item  for every state $m$ in $S$, all states $\sigma \in \fS$ 
that  are lumped into it (i.e., all elements of the fibre $\Phi_m$) 
must have the same fitness, $R_m$ (Eq.~\eref{rlump}), and
\item for every element $\sigma \in \Phi_m$, the {\em total}
mutation rate to `target types'  in $\Phi_n$, i.e.,
$\sum_{\tau \in  \Phi_n} \cM_{\sigma, \tau}$,
must be the same; it may depend on $n$ and $m$, but not on which
particular element $\sigma \in \Phi_m$ is considered.
Note, however, that only the total mutation rates are
relevant, not how they are distributed across the various
types in $\Phi_n$; see Eq.~\eref{mlump} and Fig.~1.
\end{enumerate}

Well-known examples that allow for lumping are  evolution 
models on sequence space $\varSigma^N$,
the set of possible sequences
of length $N$ over an alphabet $\varSigma$ (e.g., $\varSigma =
\{A,G,C,T\}$ or $\varSigma=\{-1,1\}$),
provided all sites
mutate independently and according to the same process,
and the fitness function is invariant under permutation of sites.
Independent mutation is a perfectly natural standard
assumption; permutation invariance of fitness is more of a restriction,
but still a common assumption.  It applies,
for example, if fitness only depends on the sequence through
the number of mutated positions (i.e., the Hamming distance)
relative to the wildtype or some other
reference sequence. Specifically, the fitness of regulatory
sequences has been modelled  as a hyperbolic function of
their binding energy to
the regulatory protein, which, in a good approximation, depends
linearly on the number of mismatches relative
to the perfectly matching sequence \cite{GeHw02}.
Then, $S=\{0,\ldots,N\}^{d}$
with $d=\lvert \varSigma \rvert$ is  the obvious choice, where
the elements of $S$ are given by
$i=(i_{\ell})_{\ell \in \varSigma}$
with $i_{\ell}$ denoting the number of sites occupied by letter $\ell$.
In fact, $d= \lvert \varSigma \rvert -1$
is also sufficient due to the constraint 
$\sum_{\ell \in \varSigma} i_{\ell} = N$.
If $\varSigma=\{-1,1\}$ and if we assume parallel mutation and selection, 
we arrive at a special case of the single-step mutation model 
in \sref{ex}. Namely, on $\varSigma^N$, the non-diagonal elements
of the mutation generator are $\cM_{\sigma,\tau} = \mu/N$ if $\sigma$ and
$\tau$ differ at exactly one site, while all other elements vanish;
on $S=\{0,1,\ldots,N\}$, we get
\begin{equation}\label{eq:u_seq}
U_i^+=\mu \frac{N-i}{N} = M_{i,i+1} \quad \text{and} \quad 
U_i^-=\mu \frac{i}{N} = M_{i,i-1}
\end{equation}
as the `lumped' mutation rates (since $N-i$ and $i$, respectively, are
the number of ways in which a sequence with $i$ mismatches may
mutate into one with $i+1$ or $i-1$ mismatches in one step).

For simple situations like this one, the above lumping according
to the Hamming distance
is routinely used, one way or another (see, e.g.,
\cite{NoSc89,GeHw02}). It is also implicit in many multilocus
models; here, the original genotype
is usually not considered at all, and one entirely relies
on some effective model as identified with the number of mutated sites
relative to some wild-- or optimal type, see \cite{Kond88,Char90}. 

With somewhat more effort, models with a nucleotide alphabet
may be treated along the same lines \cite{GaGr04}, this time,
with $d=\lvert \varSigma \rvert -1 = 3$. What is less obvious is 
that the procedure 
also works for more interesting fitness functions like those that
arise from Hopfield models. 
Here, again, $\varSigma = \{-1,1\}$, but, this time, 
fitness is assigned according to
the sitewise comparison of the sequence with {\em several}
reference sequences (known as patterns). Such fitness functions
have multiple peaks, are  tunably rugged, and  fail to be
permutation invariant across all sites. 
Rather, the set of sites $\Lambda = \{1,2,\ldots,N\}$ may be
partitioned into $d=K$ (disjoint) subsets so that the sites in each
subset are equivalent with respect to {\em all}  reference sequences.
Consequently, permutation invariance still applies within subsets, and the
effective type now is a $d$-tuple of letter frequencies, each taken over
the sites in a given subset.
For details, see \sref{applic}, and Fig.~2.

\smallskip

{\em Continuous approximation:} 
Even after lumping, the state space is usually large,
typically $S=\{0,1,\ldots, N\}^d $ with large $N$
and moderate $d$. In a second
simplification step (that may, of course, be applied independently
if the model was on $S$ in the first place), we now replace the
discrete variational problem by a continuous one on
a compact domain $D \in \RR^d$. As described in \sref{maxeval}, 
the discrete type $i \in S$ is replaced by $x_i=i/N$ in $S/N$,
and approximated  by a continuous variable $x$ 
in the limit $N \to \infty$. For the two-state model
discussed above, $x \in [0,1]$ is
simply the fraction of mutated sites relative to the reference sequence.
(In population genetics, the infinite sites limit
$N \to \infty$ at constant $i$ (and hence $i/n \to 0$) is more familiar;
for a discussion of how this relates to the limiting procedure here,
see \cite{HRWB02} and \cite{BaWa01}).
For models with a nucleotide alphabet, $x \in [0,1]^3$ tells us at
which fraction of the  sites there is a replacement of the
reference letter  by 
one of the three other nucleotides (in a suitable encoding).
Finally, in the Hopfield model, $x \in [0,1]^d$ holds the fractions
of sites that read `$+1$'  within the $d$ subsets. 

\smallskip

Our {\em main result}, Theorem \ref{theo:main},
now rephrases the variational problem in 
terms of matrices $E$ and $F$
that result from symmetrization of $M$, and hence of $M+R$.
$F$ is the symmetrized mutation matrix, as far as the
non-diagonal elements are concerned; its diagonal elements
are arranged so that $F$ is a Markov generator. $E$ is a
diagonal matrix that holds both the reproduction rates and
contributions from the mutation rates.  

Theorem \ref{theo:main} now tells us that, under certain conditions on
$E$ and $F$, a large simplification 
relative to the discrete problem
is obtained: The variational problem boils down to
a continuous one on $D \subset \RR^d$. If $d$ is small, this
can often be solved explicitly. Let us now first discuss these
conditions, and then the result, in more detail.

The assumptions on $E$ and $F$ in \eref{contapproxe}, \eref{contapproxf}
and \eref{decayfk}  appear to 
be rather special, but they are, in fact,
very natural for many models in population genetics. 
The {\em continuous approximation of the matrices $E$ and $F$}, as
imposed by \eref{contapproxe} and \eref{contapproxf}, always 
applies if  the reproduction and mutation rates have their own continuous
approximations each (i.e., $R_i = r(i/N) + \cO(1/N)$ and 
$M_{ij} = u_k(i/N) + \cO(1/N)$ with $C^2_b(D,\RR)$ functions
$r$ and $u_k$ for all $i,j$, where $k=j-i$) as in the
single-step mutation model (\sref{ex} and Eq.~\eref{u_seq}). 
For lumped versions of
sequence space models, the condition on the mutation part
is always fulfilled;
often, the continuous version is even exact, i.e.\ without
the $\cO(1/N)$ term, as we see from \eref{u_seq}. As to the 
reproduction rates, the condition requires that the fitness function
becomes locally smooth  when the types become continuous
(but this does not exclude  ruggedness at a
larger scale). Many models in population genetics rely on this
assumption, in particular, the usual models of quantitative
genetics (for review, see \cite{Buer00}).

Furthermore, $F$ must {\em decay sufficiently fast away from
the diagonal} (Eq.~\eref{decayfk}).
If we have a suitable distance between types
and mutation decays fast enough with distance, then, 
with a suitable indexing,
the symmetrized mutation matrix $F$ will be concentrated around its diagonal. 
In the single-step mutation model, $M$ is  tridiagonal, and hence
\eqref{eq:decayfk} is trivially satisfied.
In many other models (such as the random walk mutation model with
Gaussian mutant distribution), the decay is exponential
and hence even faster than the cubic decay
required in \eref{decayfk}.

Under the conditions just discussed, it turns out that the remaining
variational problem  involves only the diagonal
term $E$; $F$ contributes only an 'irrelevant' $\cO(1/N)$ term.
The maximum of the continuous function $E(x)$ that approximates the
entries of $E$ then yields  the leading eigenvalue, or mean fitness,
in leading order. 
For the single-step mutation model ($d=1$),  $E(x)$ is easily seen
to be $E(x) = r(x) - g(x)$ (cf.\ \eref{rg}), where $r$ is the (continuous
approximation of) the fitness of type $x$, and $g(x)$  has a
plausible interpretation as
loss in fitness due to mutation \cite{HRWB02}. The 
explicit expression for $E(x)$ is immediate in this case
since
the mutation matrix is tridiagonal. 
In nontrivial examples, however, more work is required
to get this function explicitly; examples will be presented
in a forthcoming paper.

In the generic case that $E(x)$ has  a unique, quadratic maximum,
we can further say  that the
{\em ancestral distribution} is concentrated around the
point $x^*$ at which $E(x)$ assumes its maximum. More precisely,
any given fraction  of at least $1-\beta$  of the
distribution's mass is contained in an interval centred
at $x^*$ whose width decreases as $1/\sqrt{N}$ (Theorem \ref{theo:a}).
As a consequence, $x^*$ obtains the interpretation of 
the mean ancestral type, up to an
error term of the order of at most $1/N^{1/3}$ (Corollary \ref{cor:mean}).

Open questions  
concern the stationary distribution in
the {\em present} population, and quantities associated
with it. In the single-step  model, the mean type of the present
population is available through the inverse function
of $r$ evaluated at $\lambdamax$ (if $r$ is monotonic);
this also leads the way to other properties of the
distribution, in particular, the variance of the present
type, and the variance in fitness \cite{HRWB02}.
This does, however, not 
carry over to higher dimensions in a simple way --
the present seems to be more difficult to deal with
than the past! For the same reason, the 
criteria for the existence of error thresholds
given in \cite{HRWB02} remain to be generalized.

The motivation  for this work came from continuous-time
mutation-repro\-duction (or mutation-selection) models (cf.\ \eref{lindgl}, 
\eref{nonlindgl} and \eref{coupled}),
which also set the scene for this discussion.
However, it should have become clear that our
results are not tied to these specific models.
Our main result
(Theorem \ref{theo:main}) simply yields asymptotic estimates for
the leading eigenvalues of large  matrices that
possess a certain  continuous approximation, and whose
elements decay sufficiently fast away from the
diagonal. These properties are shared by many dynamical
systems (in discrete and continuous time); obvious
candidates are models with migration and spatially varying
growth rate (see \cite[Chap.\ II]{Kot01} for an overview of
spatially structured population models).

\bigskip
\begin{acknowledgement}
Support and hospitality of the Erwin Schr\"odinger International Institute for
Mathematical Physics in Vienna, where this work was initiated
and partly carried out,
is gratefully acknowledged. A.B.\ also thanks the German  
Research Council (DFG) for  partial support in
the Dutch-German Bilateral Research Group  "Mathematics of random Spatial 
Models from Physics and Biology".
It is our pleasure to thank T.\ Garske and P.\ Blanchard for
helpful discussions, 
and R.\ B\"urger and an anonymous referee for  reading  the manuscript
very carefully and providing valuable suggestions for improvement. 
\end{acknowledgement}

%\bibliography{../eb}

\newcommand{\noopsort}[1]{} \newcommand{\printfirst}[2]{#1}
  \newcommand{\singleletter}[1]{#1} \newcommand{\switchargs}[2]{#2#1}

\end{document}